\documentclass[aps,pra,twocolumn,groupedaddress,nofootinbib]{revtex4-1}
\usepackage{amsmath}
\usepackage{amssymb}
\usepackage{graphicx,subfigure}
\usepackage{mathrsfs}
\usepackage[mathscr]{eucal}
\usepackage{ntheorem}
\usepackage[T1]{fontenc}
\usepackage{bm}
\usepackage{subfigure}
\usepackage[bookmarks=false]{hyperref}
\usepackage{xcolor}
\usepackage{float}

\hypersetup{colorlinks=true,citecolor=blue,
linkcolor=blue,urlcolor=blue,pdfstartview=FitH,
bookmarksopen=true}

\newcommand{\ket}[1]{|#1\rangle}
\newcommand{\bra}[1]{\langle #1|}

\newcommand{\expt}[1]{\langle #1 \rangle}

\newcommand{\tr}{\mathrm{tr}}

\newcommand{\abs}[1]{\lvert #1\rvert}
\newcommand{\norm}[1]{\lVert #1\rVert}

\def\CC{{\rm\kern.24em \vrule width.04em height1.46ex depth-.07ex \kern-.30em C}}
\def\RR{{\rm\kern.24em \vrule width.04em height1.46ex depth-.07ex
\kern-.30em R}}
\def\P{{\rm I\kern-.25em P}}
\def\GJ{\textcolor{black}}

\begin{document}

\title{Dissipative Adiabatic Measurements: Beating the Quantum Cram\'{e}r-Rao Bound}
\author{Da-Jian Zhang}
\affiliation{Department of Physics, National University of Singapore, Singapore 117542}
\author{Jiangbin Gong}
\email{phygj@nus.edu.sg}
\affiliation{Department of Physics, National University of Singapore, Singapore 117542}

\date{\today}

\begin{abstract}
It is challenged only recently that the precision attainable in any measurement of a physical parameter is fundamentally limited by the quantum Cram\'{e}r-Rao Bound (QCRB). Here, targeting at measuring parameters in strongly dissipative systems, we propose an innovative measurement scheme called {\it dissipative adiabatic measurement} and theoretically show that it can beat the QCRB.  Unlike projective measurements, our measurement scheme, though consuming more time,  does not collapse the measured state and,
more importantly, yields the expectation value of an observable as its measurement outcome, which is directly connected to the parameter of
interest. Such a direct connection \GJ{allows to extract} the value of the parameter from the measurement outcomes in a straightforward
manner, with no fundamental limitation on precision in principle.  Our findings not only provide a marked insight into quantum metrology but also are highly useful in dissipative quantum information processing.
\end{abstract}

\maketitle

\section{Introduction}

Improving precision of quantum measurements underlies both technological and scientific progress. Yet, it has never been doubted until recently \cite{2017Seveso12111} that the precision attainable in any quantum measurement of a physical parameter is bounded by the inverse of the quantum Fisher information (QFI) multiplied by the number of the measurements in use \cite{1976Helstrom}. Such a bound, known as the celebrated
quantum Cram\'{e}r-Rao Bound (QCRB), has been deemed as the ultimate precision allowed by quantum mechanics that cannot be surpassed under any circumstances \cite{1994Braunstein3439,2009Paris125}. As such, ever since its inception in 1967 \cite{1967Helstrom101}, the QCRB has been the cornerstone of quantum estimation theory underpinning virtually all aspects of research in quantum metrology \cite{2004Giovannetti1330,2006Giovannetti10401,2011Giovannetti222,
2018Braun35006}.

\begin{figure}[htbp]
\includegraphics[width=0.44\textwidth]{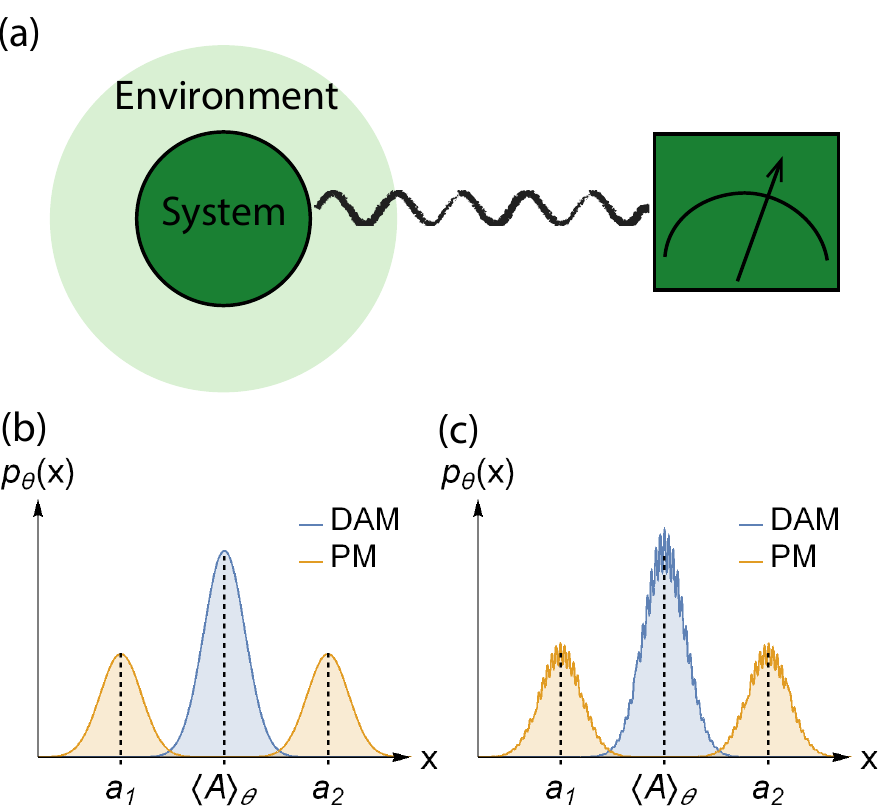}
\caption{Schematic diagrams. (a) The setup: a dissipative system coupled to a measuring apparatus. The coupling is extremely strong but instantaneous for PMs whereas it is very weak but of long duration for DAMs. (b) An ideal PM corresponds to the strong coupling and short-time limit, with eigenvalues $a_i$ of the measured observable $A$ as its outcomes.  In contrast, an ideal DAM corresponds to the weak coupling and long-time limit, with the expectation value $\expt{A}_\theta$ as its outcome. \GJ{Outcomes from both DAMs and PMs in their ideal cases are still subject to fluctuations (represented by the shaded spread in the plotted probability distribution) originating from position uncertainties in the initial state of the apparatus.} (c) In practice, the coupling strength and coupling time are finite, causing some deviations from ideal cases for both PMs and DAMs.}
\label{fig1}
\end{figure}

In this work, inspired by Aharonov \textit{et al.}'s adiabatic measurements \cite{1993Aharonov38}, we propose an innovative measurement scheme tailored for strongly dissipative systems.  Measurements based on our scheme are coined ``dissipative adiabatic measurements'' (DAMs). The system to be measured here is a strongly dissipative system initially prepared in its steady state and then coupled to a measuring apparatus via an extremely weak but long-time interaction (see Fig.~\ref{fig1}a). The dynamics of the coupling procedure is dominated by the dissipative process, which continuously projects the system into the steady state. Such a dissipation-induced ``quantum Zeno effect'' \cite{2000Beige1762} effectively decouples the system from the apparatus in the long time limit, suppressing the so-called quantum back action of measurements \cite{2004Giovannetti1330}. Unlike projective measurements (PMs), a DAM therefore does not collapse the measured state, namely, the steady state. Moreover, as shown below, its outcome is the expectation value of the measured observable in the steady state, \GJ{up to some fluctuations due to position uncertainties in the initial state of the apparatus}.

Aided by a simple yet well known model, we show that the QCRB can be beaten by DAMs although it does hold for (ideal) PMs. So, despite \GJ{the long time requirement in DAMs}, their ability of beating the QCRB comes as a surprise and constitutes a significant advancement in our understanding of quantum metrology. Indeed, the outcomes of DAMs are directly connected to some parameter of interest whereas this is never the case for PMs. Such a direct connection allows \GJ{to extract} the value of the parameter from the outcomes in a straightforward
manner, \GJ{in principle without} any fundamental limitation on precision. To solve the conflict between our results and the QCRB, we resort to Ref. \cite{2017Seveso12111} which shows that the QCRB relies on a previously overlooked assumption. This assumption can be severely violated by DAMs. Our work therefore provides an intriguing measurement scheme that can surpass the so-called  ``ultimate precision'' limit.

Apart from being of fundamental interest, our results are important from an application perspective as well. On one hand, DAMs allow for detecting steady states without perturbing them. This somewhat exotic feature could be highly useful in dissipative quantum information processing  \cite{1999Plenio2468,2008Diehl878,2009Verstraete633,2010Weimer382,2011Kastoryano90502,
2011Cho20504,2011Krauter80503,2011Barreiro486,
2011Vollbrecht120502,2013Kastoryano110501,2013Carr33607,2013Torre120402,2013Rao33606,
2014Bentley40501,2016Zhang12117,2016Zhang52132,2016Abdi233604,
2016Kimchi-Schwartz240503,2016Reiter40501,
2016Znidaric30403,2017Reiter1822}, where steady states are typically entangled states or some other desirable states. On the other hand, as shown below, decoherence and dissipation play an integral part in DAMs.
As a consequence, if the effects of decoherence and dissipation become increasingly strong, the efficiency of DAMs would increase instead of declining. This theoretical view, verified by numerical
simulations below, suggests the fascinating possibility that decoherence and dissipation may be exploited for good rather than causing detrimental effects in quantum measurements.

\section{preliminaries}

We start with some preliminaries. Given a family of quantum states $\rho_\theta$ characterized by an unknown parameter $\theta$, a basic task in quantum metrology is to estimate $\theta$ as precisely as possible by using $N$ repeated measurements. Any measurement can be described by a positive operator-valued measure (POVM) $\Pi_x$ satisfying $\int dx\ \Pi_x=\openone$, with $x$ labeling its outcome. To obtain an estimate of $\theta$, one can input the outcomes $x_1,\cdots,x_N$ of $N$ measurements into an estimator $\hat{\theta}(x_1,\cdots,x_N)$, which is a map from the set of outcomes to the parameter space. Upon
optimizing over all unbiased estimators for a given measurement, one reaches the classical Cram\'{e}r-Rao bound (CCRB) $\textrm{Var}(\hat{\theta})\geq[NF(\theta)]^{-1}$. Here $\textrm{Var}(\hat{\theta})$ is the variance of $\hat{\theta}$ and $F(\theta)=\int dxp_\theta(x)[\partial_\theta\ln p_\theta(x)]^2$ is the classical Fisher information (CFI), with $p_\theta(x)=\tr(\Pi_x\rho_\theta)$ denoting the conditional probability density of getting outcome $x$ when the actual value of the parameter is $\theta$. To find the best precision, one needs to further optimize the CFI $F(\theta)$ over all possible measurements $\Pi_x$.
In doing this, it has been implicitly assumed that $\Pi_x$ is independent of $\theta$ \cite{1994Braunstein3439,2009Paris125,1967Helstrom101}. Under this assumption, referred to as the $\theta$-independence assumption hereafter, one can prove that $F(\theta)\leq H(\theta)$, with $H(\theta)=\tr(\rho_\theta L_\theta^2)$ denoting the QFI expressed in terms of the symmetric logarithmic derivative $L_\theta$, i.e., the Hermitian operator satisfying $\partial_\theta\rho_\theta=\frac{1}{2}(L_\theta\rho_\theta+\rho_\theta L_\theta)$. Substituting $F(\theta)\leq H(\theta)$ into the CCRB, one arrives at the QCRB $\textrm{Var}(\hat{\theta})\geq[NH(\theta)]^{-1}$. As the $\theta$-independence assumption has never been doubted until the recent work \cite{2017Seveso12111}, it has been believed for a long time that the QCRB is universally valid for all kinds of measurements and, therefore, represents the ultimate precision allowed by quantum mechanics. Moreover, the optimal POVM (asymptotically) saturating this bound has been shown to be the PM associated with the observable $A_\textrm{opt}=\theta\openone+L_\theta/H(\theta)$ \cite{2009Paris125}. The proof of the QCRB is revisited in Appendix \ref{app:A}. Shortly, we will show that DAMs violate the $\theta$-independence assumption and can be exploited to beat the QCRB.

To better digest the above knowledge, we may take the generalized amplitude damping process \cite{2010Nielsen} as an example. It is described by the Lindblad equation, $\frac{d}{d t}\rho(t)=\mathcal{L}_\theta\rho(t)$, where
$\mathcal{L}_\theta\rho:=\gamma[
\theta(\sigma_{-}\rho\sigma_{+}-\frac{1}{2}\{\sigma_{+}
\sigma_{-},\rho\})
+(1-\theta)(\sigma_{+}\rho\sigma_{-}-\frac{1}{2}\{\sigma_{-}
\sigma_{+},\rho\})]$ is a Liouvillian superoperator depending on the parameter $\theta\in(0,1)$, with $\gamma$ denoting the decay rate, $\sigma_-=\ket{0}\bra{1}$, and $\sigma_+=\ket{1}\bra{0}$.
For this dissipative process, there is a unique steady state  $\rho_\theta=\textrm{diag}(\theta,1-\theta)$, which can be approached if the evolution time is much longer than the relaxation time of the process. Our purpose here is to infer the value of $\theta$ from $N$ repeated measurements on $\rho_\theta$. Considering that $\theta$ is a monotone function
of the temperature of environment \cite{2010Nielsen}, estimation of $\theta$ offers one means of temperature estimation that has received much attention recently in quantum thermometry \cite{2017Mancino130502,2019Mehboudi30403,2019Seah180602}.
 Using $\rho_\theta=\textrm{diag}(\theta,1-\theta)$, we have
$L_\theta=\textrm{diag}(\frac{1}{\theta},-\frac{1}{1-\theta})$ and further obtain $H(\theta)=\frac{1}{\theta(1-\theta)}$. So, the QCRB reads $\textrm{Var}(\hat{\theta})\geq \theta(1-\theta)/N$. On the other hand, direct calculations show that $A_\textrm{opt}=\ket{0}\bra{0}$. Clearly, the PM associated with $A_\textrm{opt}$ has two potential outcomes $1$ and $0$, with the probabilities $\theta$ and $1-\theta$, respectively. Intuitively, one may think of $1$ and $0$ as the head and tail of a coin, with $\theta$ corresponding to the coin's propensity to land heads. Performing $N$ such PMs, we can obtain a series of data, $x_1^\textrm{PM},\cdots,x_N^\textrm{PM}$, with $x_i^{\textrm{PM}}\in\{1,0\}$. Then the value of $\theta$ can be estimated as,
$\hat{\theta}(x_1^\textrm{PM},\cdots,x_N^\textrm{PM}):=
\sum_{i=1}^Nx_i^{\textrm{PM}}/N$, amounting to the frequency of $1$ appearing in the data. Using well known results regarding $N$-trial coin flip experiments, we have
\begin{eqnarray}\label{Ex-error}
\textrm{Var}(\hat{\theta})=\theta(1-\theta)/N,
\end{eqnarray}
indicating that the PM associated with $A_\textrm{opt}$ already reaches the QCRB. In passing, given the same amount of resources, e.g., the number of measurements, one may wonder whether the estimation error can be further reduced, e.g., by measuring a (possibly entangled) state other than $\rho_\theta$. In Appendix \ref{app:B}, we prove that the answer is negative if the QCRB is valid.
In order to simplify our paper as much as we can, throughout this paper, we use the above simple example to demonstrate our findings, although our findings are generally applicable to dissipative systems.

\section{Dissipative adiabatic measurement}

Keeping this example in mind, we proceed to develop DAMs. Suppose that we are given a dissipative system $\mathscr{S}$ with a Liouvillian superoperator $\mathcal{L}_\theta$. Here we take $\theta$ to be a single parameter for simplicity. $\mathcal{L}_\theta$ is assumed to be such that: (a) it admits a unique steady state $\rho_\theta$; (b) the nonzero eigenvalues $\lambda_h$ ($h>0$) of $\mathcal{L}_\theta$ have negative real parts, that is, there is a dissipative gap $\Delta:=\min_{h>0}\abs{\textrm{Re}(\lambda_h)}$ in the Liouvillian spectrum. $\mathscr{S}$ is initially prepared in its steady state $\rho_\theta$. Note that this is achievable even though $\theta$ is unknown, as $\mathscr{S}$ automatically approaches $\rho_\theta$ because of the dissipative gap \cite{1999Plenio2468,2008Diehl878,2009Verstraete633,2010Weimer382,
2011Kastoryano90502,
2011Cho20504,2011Krauter80503,2011Barreiro486,
2011Vollbrecht120502,2013Kastoryano110501,2013Carr33607,2013Torre120402,
2013Rao33606,
2014Bentley40501,2016Zhang12117,2016Zhang52132,2016Abdi233604,
2016Kimchi-Schwartz240503,2016Reiter40501,
2016Znidaric30403,2017Reiter1822}. To simplify our discussion, we adopt
the minimal model of measurement that has been used time and again
in the literature (see Appendix \ref{app:C} for more details).
That is,
to measure an observable $A$, we add an interaction term,
$H_I=T^{-1}A\otimes \hat{p}$,
coupling $\mathscr{S}$ to a measuring apparatus $\mathscr{A}$, with coordinate and momentum denoted by $\hat{x}$ and $\hat{p}$, respectively.
Here, $T$ is a positive real number, which, in the spirit of the adiabatic theorem, will be eventually sent to infinity \cite{1993Aharonov38}. As usual, we are not interested in the dynamics of the apparatus itself and assume its free Hamiltonian to be zero \cite{1993Aharonov38}. In real situations, this could be achieved by switching to a rotating frame. The dynamics of the coupling procedure is described by the equation ($\hbar=1$),
\begin{eqnarray}\label{eq:measuring}
\frac{d}{dt}\rho(t)=\mathcal{L}_\theta
\rho(t)-i\left[H_I,\rho(t)\right]=:\mathcal{L}\rho(t).
\end{eqnarray}
If the coupling time is $T$ (so that the product of the coupling strength, i.e., $1/T$, and the coupling time is unity), $\mathscr{S}+\mathscr{A}$ undergoes the dynamical map, $\mathcal{E}_T:=e^{\mathcal{L}T}$, transforming the initial state $\rho_\theta\otimes\ket{\phi}\bra{\phi}$ to the state $\mathcal{E}_T(\rho_\theta\otimes\ket{\phi}\bra{\phi})$ at time $T$. Here, $\ket{\phi}$ denotes the initial state of $\mathscr{A}$, which is set to be a Gaussian centered at $x=0$.
After the coupling procedure, the coordinate $\hat{x}$ is observed, in order to determine the reading of the pointer. The above minimal model can be implemented in a number of experimental setups, such as in cavity quantum electrodynamics \cite{2000Hood1447,2002Mabuchi1372} and circuit quantum electrodynamics \cite{2004Blais62320,2004Wallraff162}.

To figure out the effect of $\mathcal{E}_T$, we make use of the fact
$\mathcal{L}(\rho\otimes\ket{p}\bra{p^\prime})=
(\mathcal{L}_{p,p^\prime}\rho)\otimes\ket{p}\bra{p^\prime}$, with $\mathcal{L}_{p,p^\prime}\rho:=\mathcal{L}_\theta\rho-iT^{-1}(pA\rho-p^\prime \rho A)$. Here, $\ket{p}$ denotes the eigenstate of $\hat{p}$, i.e., $\hat{p}\ket{p}=p\ket{p}$. This leads to
$\mathcal{E}_T(\rho_\theta\otimes\ket{p}\bra{p^\prime})
=(e^{\mathcal{L}_{p,p^\prime}T}\rho_\theta)\otimes\ket{p}\bra{p^\prime}$.
Note that a technical result in Ref.~\cite{2014Zanardi240406} is
\begin{eqnarray}\label{Zanardi}
\norm{e^{\mathcal{L}_{p,p^\prime}T}\mathcal{P}_\theta
-e^{\widetilde{\mathcal{L}}_{p,p^\prime}T}\mathcal{P}_\theta}=O(1/T),
\end{eqnarray}
indicating that $e^{\mathcal{L}_{p,p^\prime}T}\mathcal{P}_\theta$ gets closer and closer to $e^{\widetilde{\mathcal{L}}_{p,p^\prime}T}\mathcal{P}_\theta$ as $T$ approaches infinity \cite{1note}. Here, $\widetilde{\mathcal{L}}_{p,p^\prime}:=\mathcal{P}_\theta
\mathcal{L}_{p,p^\prime}\mathcal{P}_\theta$, and $\mathcal{P}_\theta$ denotes the projection, $\mathcal{P}_\theta(X):=(\tr_\mathscr{S}X)\rho_\theta$, mapping an arbitrary operator $X$ into $\rho_\theta$. Using Eq.~(\ref{Zanardi}) and noting that the explicit expression of $\widetilde{\mathcal{L}}_{p,p^\prime}$ reads
$\widetilde{\mathcal{L}}_{p,p^\prime}=-iT^{-1}(p-p^\prime)\expt{A}_\theta
\mathcal{P}_\theta$, where $\expt{A}_\theta:=\tr(A\rho_\theta)$, we obtain
\begin{eqnarray}\label{eq:ES}
\lim_{T\rightarrow\infty}\mathcal{E}_T(\rho_\theta\otimes\ket{p}\bra{p^\prime})
=\rho_\theta\otimes e^{-i(p-p^\prime)\expt{A}_\theta}\ket{p}\bra{p^\prime}.
\end{eqnarray}
Now, expressing $\ket{\phi}$ in $\mathcal{E}_T(\rho_\theta\otimes
\ket{\phi}\bra{\phi})$ as $\ket{\phi}=\int\phi(p)\ket{p}dp$ and using the linearity of the map $\mathcal{E}_T$ as well as Eq.~(\ref{eq:ES}), we arrive at the main formula of this paper,
\begin{eqnarray}\label{main-formula}
\lim_{T\rightarrow\infty}\mathcal{E}_T(\rho_\theta\otimes
\ket{\phi}\bra{\phi})=
\rho_\theta\otimes e^{-i\expt{A}_\theta\hat{p}}\ket{\phi}\bra{\phi}
e^{i\expt{A}_\theta\hat{p}}.
\end{eqnarray}
Formula (\ref{main-formula})
shows that in the weak coupling and long time limit, the steady state $\rho_\theta$ does not collapse and the pointer is shifted by the expectation value rather than eigenvalues of $A$ (see Fig.~\ref{fig1}b).

To gain physical insight into the above result, we compare DAMs with PMs. Both DAMs and PMs utilize interaction terms of the form, $H_I=g(t)A\otimes\hat{p}$, with $g(t)$ normalized to $\int g(t)dt=1$. Note that we have let $g(t)=1/T$ for simplicity. In PMs, this term is impulsive, that is, $g(t)=1/T$ takes an extremely large value but only for a very short time interval. So, the dominating term in Eq.~(\ref{eq:measuring}) is $H_I$. In the strong coupling and short time limit, i.e., $T\rightarrow 0$, the associated dynamical map reads
\begin{eqnarray}\label{map-pm}
\lim_{T\rightarrow 0}\mathcal{E}_T(\rho_\theta\otimes\ket{\phi}\bra{\phi})=e^{-iA\otimes\hat{p}}
\rho_\theta\otimes\ket{\phi}\bra{\phi}e^{iA\otimes\hat{p}}.
\end{eqnarray}
Clearly, this map creates correlations between
$\mathscr{S}$ and $\mathscr{A}$, giving rise to the quantum back action that the configuration of $\mathscr{S}$ after the measurement is determined by the outcome of $\mathscr{A}$.
Contrary to PMs, DAMs exploit the opposite limit of an extremely weak but long time interaction, i.e., $T\rightarrow \infty$. For this, the dissipative term $\mathcal{L}_\theta$ dominates the interaction term $H_I$ in Eq.~(\ref{eq:measuring}). The former effectively eliminates correlations created by the latter through continuously projecting $\mathscr{S}$ into its steady state $\rho_\theta$. Resulted from this nontrivial interplay is the decoupling of $\mathscr{S}$ and $\mathscr{A}$ in the long time limit, which inhibits the state of $\mathscr{S}$ from any change or collapse. Such a mechanism is suggestive of the quantum Zeno effect, making DAMs distinct from PMs in nature.

Now let us turn to the practical situation with the coupling strength and the coupling time being finite. In a PM, under the influence of $\mathcal{L}_\theta$, i.e., decoherence and dissipation, the evolved state $\mathcal{E}_T(\rho_\theta\otimes\ket{\phi}\bra{\phi})$ deviates from the ideal \textit{correlated} state given by Eq.~(\ref{map-pm}). To ensure that such deviations are small enough, it has to be required that $H_I$ is sufficiently strong so that decoherence and dissipation are comparatively weak. Evidently, such a strong interaction requirement would be increasingly difficult to meet if decoherence and dissipation become increasingly strong. This is consistent with our intuition that decoherence and dissipation are detrimental in quantum measurements. Contrary to this understanding, DAMs appreciate decoherence and dissipation as useful resources \GJ{because they are responsible for eliminating the system-apparatus correlations. Certainly, in a non-ideal DAM due to a finite $T$}, the evolved state also deviates from the ideal \textit{decoupled} state in Eq.~(\ref{main-formula}) (see Fig.~\ref{fig1}c). \GJ{Nevertheless}, the stronger decoherence and dissipation are, the more efficient the \GJ{correlation elimination} is and, therefore, the shorter the coupling time required to execute a  DAM. That is, unlike the strong interaction requirement in PMs, the long time requirement in DAMs would be increasingly easy to meet if decoherence and dissipation are increasingly strong. This point can be placed on more solid ground by virtue of perturbation theory \cite{2019Zhang}.

To illustrate the above point, we come back to the foregoing example. Here, $\gamma$ represents the strength of decoherence and dissipation; more precisely, the dissipative gap $\Delta=\gamma/2$. Note that the deviations in PM and DAM from their ideal results can be quantified by the measures
\begin{eqnarray}\label{measure-PM}
\norm{\mathcal{E}_T(\rho_\theta\otimes\ket{\phi}\bra{\phi})-
\lim_{T\rightarrow 0}\mathcal{E}_T(\rho_\theta\otimes\ket{\phi}\bra{\phi})}
\end{eqnarray}
and
\begin{eqnarray}\label{measure-DAM}
\norm{\mathcal{E}_T(\rho_\theta\otimes\ket{\phi}\bra{\phi})-
\lim_{T\rightarrow \infty}\mathcal{E}_T(\rho_\theta\otimes\ket{\phi}\bra{\phi})},
\end{eqnarray}
respectively.
\begin{figure}
\centering
\subfigure{\label{fig2a}
\includegraphics[width=.22\textwidth]{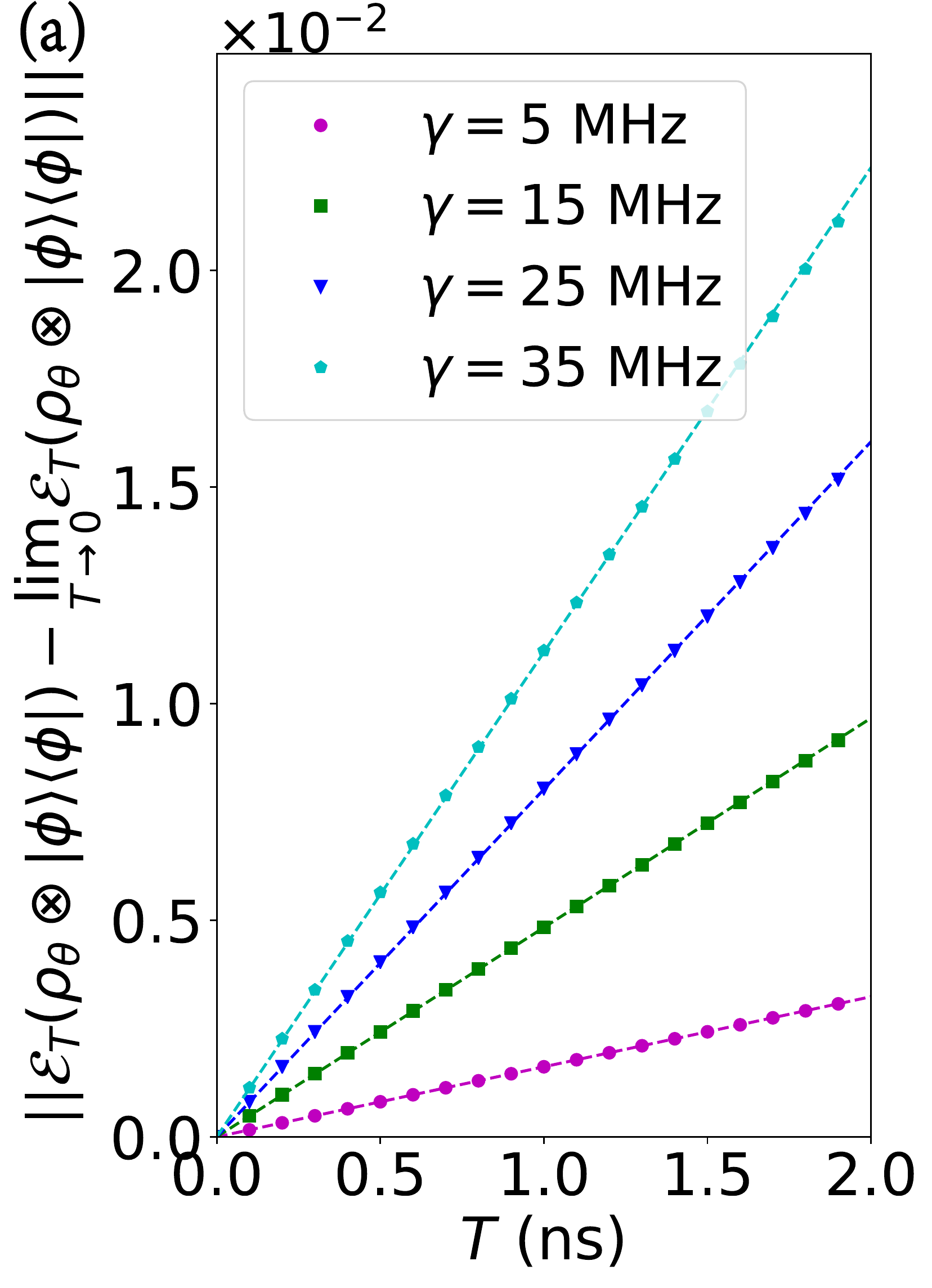}}
\subfigure{\label{fig2b}
\includegraphics[width=.22\textwidth]{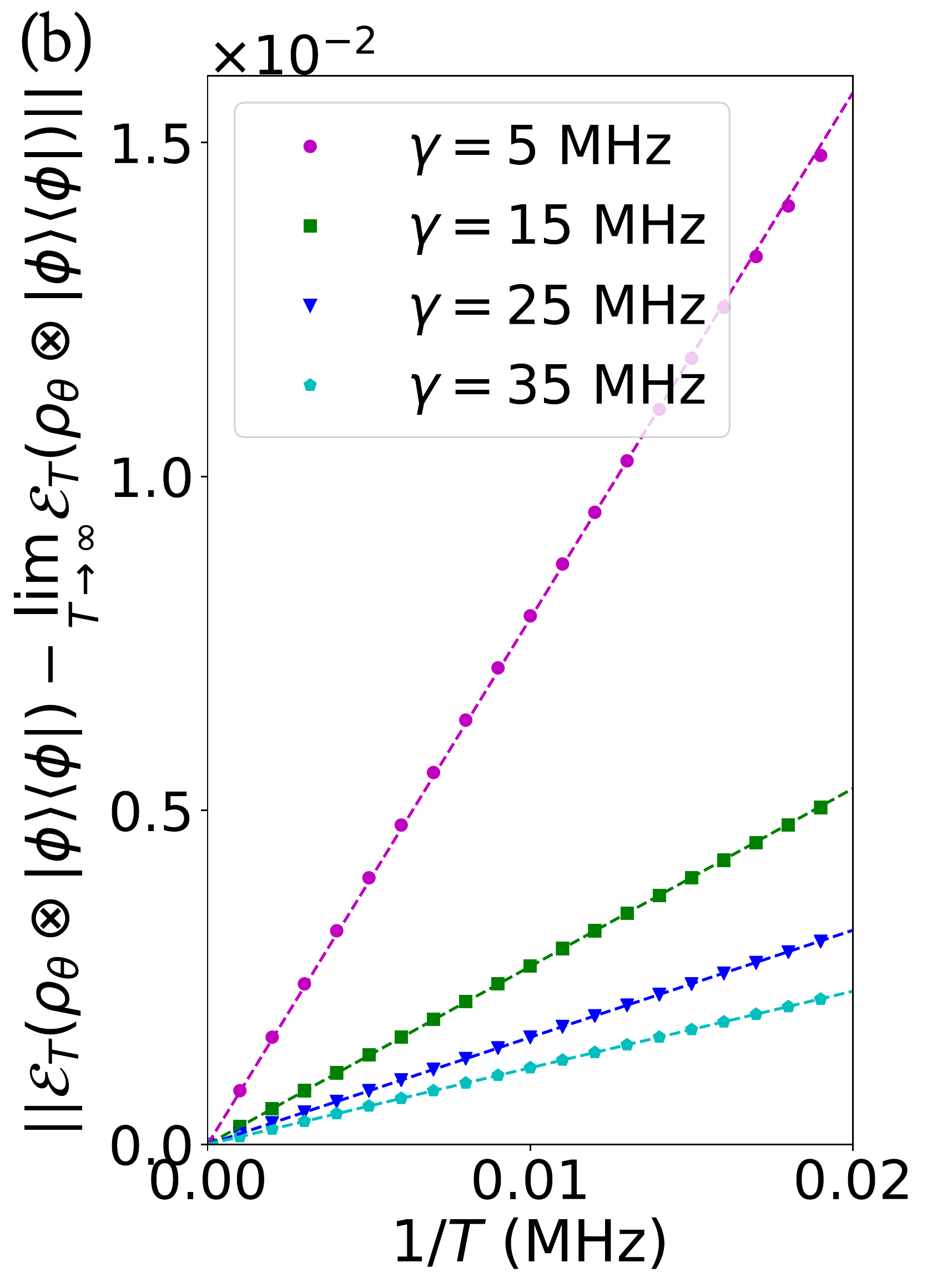}}
\caption{Numerical results of the measures in Eqs.~(\ref{measure-PM}) and (\ref{measure-DAM}) \color{black}{for (a) PM and (b) DAM with different $\gamma$.}}\label{fig2}
\end{figure}
Figure \ref{fig2} shows these two measures for different $\gamma$, where we set, without loss of generality, $A=A_\textrm{opt}=\ket{0}\bra{0}$, $\theta=1/2$, and $\sigma=1/5$, with $\sigma$ denoting the standard deviation of the Gaussian $\ket{\phi}$. \GJ{The characteristic scales of the physical quantities used in Fig.~\ref{fig2} are chosen with a realistic model in mind \cite{2013Hatridge178}}. As can be easily seen from Fig.~\ref{fig2a}, the slope of Eq.~(\ref{measure-PM}) as a function of $T$ increases as $\gamma$ increases. This indicates that $1/T$ has to be increasingly large in PM for achieving a certain desired tolerance of the deviations. In contrast, as shown in Fig.~\ref{fig2b}, the slope of Eq.~(\ref{measure-DAM}) as a function of $1/T$ decreases as $\gamma$ increases, implying that $T$ can be increasingly small in DAM for achieving the same tolerance. For instance, if the desired tolerance is set to be 0.01, $1/T$ in PM are 162, 483, 802, 1119 MHz whereas $T$ in DAM are 78.7, 26.6, 16.0, 11.5 $\mu$s, for
$\gamma=5,15,25,35$ MHz, respectively. More numerical results revealing effects arising from an intermediate coupling time are presented in Appendix \ref{app:D}.
Besides, we have examined one of the experimental setups \cite{2013Hatridge178} used to implement the minimal model of measurement adopted here and obtained analogous results (see Appendix \ref{app:E}).

\section{Beating the Quantum Cram\'{e}r-Rao Bound}

Having proposed DAMs, we now show that they can beat the QCRB. Imagine that the PM used in the foregoing example is replaced by the DAM associated with $A_\textrm{opt}=\ket{0}\bra{0}$. Analogous to $N$ PMs, $N$ such DAMs produce a series of data, $x_i^\textrm{DAM}$, $i=1,\cdots,N$, where $x_i^\textrm{DAM}$ denotes the outcome of the $i$th DAM. As before, these data can be further input into the estimator, $\hat{\theta}(x_1^\textrm{DAM},\cdots,x_N^\textrm{DAM})=\sum_{i=1}^N x_i^\textrm{DAM}/N$, to obtain an estimate of $\theta$. Note that the probability density of getting $x_i^\textrm{DAM}$ in the (ideal) DAM is
\begin{eqnarray}\label{pro-DAM}
p_\theta(x_i^\textrm{DAM})=\frac{1}{\sqrt{2\pi\sigma^2}}
e^{-\frac{(x_i^\textrm{DAM}
-\theta)^2}{2\sigma^2}},
\end{eqnarray}
which can be verified by using Eq.~(\ref{main-formula}) and noting that $\expt{A_\textrm{opt}}_\theta=\theta$ (see Appendix \ref{app:F}).
It follows that $\hat{\theta}(x_1^\textrm{DAM},\cdots,x_N^\textrm{DAM})$ is an unbiased estimator with
\begin{eqnarray}\label{Ex-DAM-error}
\textrm{Var}(\hat{\theta})=\sigma^2/N.
\end{eqnarray}

For the sake of comparing Eq.~(\ref{Ex-DAM-error}) with Eq.~(\ref{Ex-error}), we stress that Eq.~(\ref{Ex-error}) is, strictly speaking, obtained under the limiting condition that the PM is ideal and $\sigma\rightarrow 0$. Indeed, the probability density of getting $x_i^\textrm{PM}$ in the (ideal) PM is (see Appendix \ref{app:F})
\begin{eqnarray}\label{pro-PM}
p_\theta(x_i^\textrm{PM})=\frac{\theta}{\sqrt{2\pi\sigma^2}}
e^{-\frac{(x_i^\textrm{PM}-1)^2}{2\sigma^2}}+
\frac{1-\theta}{\sqrt{2\pi\sigma^2}}e^{-\frac{{x_i^\textrm{PM}}^2}{2\sigma^2}}.
\end{eqnarray}
Equation (\ref{pro-PM}) is, as a matter of fact, a continuous probability distribution but can be treated as the discrete probability distribution giving rise to Eq.~(\ref{Ex-error}) in the limit of $\sigma\rightarrow 0$ (see Appendix \ref{app:F}). In contrast to Eq.~(\ref{Ex-error}), $\textrm{Var}(\hat{\theta})$ in Eq.~(\ref{Ex-DAM-error}) approaches zero as $\sigma\rightarrow 0$, indicating that there is no fundamental limitation on precision in the DAM. Here we point out that $\textrm{Var}(\hat{\theta})$ in Eq.~(\ref{Ex-DAM-error}) is, as a matter of fact, limited by the the standard deviation of
the normal distribution (\ref{pro-DAM}), but this is not a fundamental limitation but a limitation stemming from non-ideal experimental conditions. Furthermore, it can be shown that the POVM operator associated with the DAM is  $\Pi_{x^\textrm{DAM}}=
\frac{1}{\sqrt{2\pi\sigma^2}}e^{-\frac{(x^\textrm{DAM}-\theta)^2}{2\sigma^2}}
\openone$ (see Appendix \ref{app:F}). Evidently, it violates the $\theta$-independence assumption, which is the technical reason for the DAM to be able to beat the QCRB.

Lastly, it may be instructive to give a simple physical picture for comprehending the result that the QCRB can be beaten by the DAM. To do this, suppose that we are given only one data, $x_i^\textrm{PM}$ or $x_i^\textrm{DAM}$. Roughly speaking, $x_i^\textrm{PM}$ is either $1$ or $0$, which is unrelated to $\theta$. Thus, there is no way for us to accurately infer $\theta$ from $x_i^\textrm{PM}$. However, for a small $\sigma$, $x_i^\textrm{DAM}$  takes a value equal to or close to $\expt{A_\textrm{opt}}_\theta=\theta$, which is directly related to $\theta$. Based on this direct relationship, one can obtain a fairly good estimate from $x_i^\textrm{DAM}$. In this sense, $x_i^\textrm{DAM}$ is more informative than $x_i^\textrm{PM}$. To further confirm this point, one can work out their CFI. The CFI associated with Eq.~(\ref{pro-DAM}) reads $F^\textrm{DAM}(\theta)=1/\sigma^2$, whereas that associated with Eq.~(\ref{pro-PM}) satisfies $F^\textrm{PM}(\theta)\leq H(\theta)$, no matter how small $\sigma$ is (see Appendix \ref{app:F}).  That is, $F^\textrm{PM}(\theta)$ is bounded by the QFI but $F^\textrm{DAM}(\theta)$ is unbounded in principle.

\section{Concluding remarks}

Before concluding, we make a few remarks. Equation (\ref{Ex-DAM-error}) is nothing but the CCRB for the normal distribution (\ref{pro-DAM}). Our finding here is that the CCRB (\ref{Ex-DAM-error}) can be infinitely small in principle and, in particular, can be smaller than the QCRB (\ref{Ex-error}). This clearly demonstrates that there is no intrinsic limitation on precision, which is contrary to the quantum Cram\'{e}r-Rao theorem \cite{1994Braunstein3439,2009Paris125}. On the other hand, the steady state in our example is a classical state and, therefore, does not display any coherent nature. We emphasize that the reason of choosing this example is due to its simplicity. Our measurement is certainly capable of revealing coherent nature of a quantum system as long as the steady state in question is genuinely quantum. Any system studied in Refs.~\cite{2009Verstraete633,2011Kastoryano90502,
		2011Cho20504,2011Krauter80503,2013Carr33607,2013Torre120402,2013Rao33606,
		2014Bentley40501,2016Abdi233604,
		2016Kimchi-Schwartz240503,2016Reiter40501,
		2016Znidaric30403}, as far as we can see, may serve as one illustrative example.

In conclusion, we have proposed an innovative measurement tailored for strongly dissipative systems. In our measurement, the dissipative dynamics continuously eliminates correlations created by the weak system-apparatus interaction, resulting in the decoupling of the system from the apparatus in the long time limit. Unlike PMs, our measurement therefore does not collapse the measured state, and moreover, its outcome is the expectation value of an observable in the steady state, which is directly connected to the parameter of interest. By virtue of this, our measurement is able to beat the QCRB, without suffering from any fundamental limitation on precision. These findings, solidified by a simple but delightful example, provide a revolutionary insight into quantum metrology. We highlight that our measurement works in a state-protective fashion and embraces decoherence and dissipation as useful resources. Such exotic features could be highly useful in dissipative quantum
information processing.

\begin{acknowledgments}
D.-J.Z.~thanks Dianmin Tong, Ya Xiao, Xizheng Zhang, and Sen Mu for helpful discussions.
J.G.~is supported by the
Singapore NRF Grant No.~NRF-NRFI2017-04 (WBS No.~R-144-000-378-281).
D.-J.Z.~acknowledges support from the National Natural Science Foundation of
China through Grant No.~11705105 before he joined NUS.
\end{acknowledgments}

\appendix
\setcounter{equation}{0}

\section{Revisiting the proof of the QCRB} \label{app:A}

Here, focusing on the single-parameter case,
we revisit the proof of the QCRB
\cite{1967Helstrom101,1994Braunstein3439,2009Paris125}.
Suppose that one is given many copies of a quantum state $\rho_\theta$
characterized by an unknown parameter $\theta$, and \GJ{wishes} to estimate $\theta$
as precisely as possible by using $N$ repeated measurements.
In general, a quantum measurement can be described by
a POVM $\{\Pi_x\}$. Here, $\Pi_x$ is a positive-semidefinite operator
satisfying $\int dx\Pi_x=\openone$, with $\openone$ denoting the identity
operator. $x$ labels the ``results'' of the measurement.
Although written here as a single continuous real variable, $x$ can be discrete or
multivariate. Accordingly, the outcomes of $N$ repeated measurements can be
expressed as $x_1,x_2,\cdots,x_N$. In order to extract the value of $\theta$ from these data, one can resort to an estimator,
$\hat{\theta}(x_1,\cdots,x_N)$, which is a map
from the data $x_1,\cdots,x_N$ to the parameter space.

Given a set of outcomes $x_1,\cdots,x_N$, one can obtain an estimate $\hat{\theta}(x_1,\cdots,x_N)$ of $\theta$. The deviation of the estimate from the true value of $\theta$ can be quantified by
\cite{1994Braunstein3439}
\begin{eqnarray}\label{df:Delta-theta}
\delta\theta(x_1,\cdots,x_N):=\frac{\hat{\theta}(x_1,\cdots,x_N)}
{\abs{d\expt{\hat{\theta}}/d\theta}}
-\theta.
\end{eqnarray}
Here, $\expt{\hat{\theta}}$ denotes the statistical average
of $\hat{\theta}(x_1,\cdots,x_N)$ over potential outcomes $x_1,\cdots,x_N$,
\begin{eqnarray}
\expt{\hat{\theta}}=
\int dx_1\cdots dx_N\,p_\theta(x_1)
\cdots p_\theta(x_N)\hat{\theta}(x_1,\cdots,x_N),\nonumber\\
\end{eqnarray}
where
\begin{eqnarray}\label{df:p}
p_\theta(x)=\tr(\Pi_x\rho_\theta).
\end{eqnarray}
The derivative $d\expt{\hat{\theta}}/d\theta$ appearing
in Eq.~(\ref{df:Delta-theta}) is used to remove the local difference
in the ``units'' of the estimate and the parameter \cite{1994Braunstein3439}.
The estimation error can be then defined as \cite{1994Braunstein3439}
\begin{eqnarray}
{\expt{(\delta\theta)^2}},
\end{eqnarray}
i.e., the statistical average
of $\delta\theta^2(x_1,\cdots,x_N)$ over potential outcomes
$x_1,\cdots,x_N$. In particular, if $\hat{\theta}(x_1,\cdots,x_N)$
is unbiased, i.e., $\expt{\hat{\theta}}=\theta$, there is
\begin{eqnarray}
\delta\theta(x_1,\cdots,x_N)=\hat{\theta}(x_1,\cdots,x_N)-\expt{\hat{\theta}},
\end{eqnarray}
indicating that $\expt{(\delta\theta)^2}$ is simply the variance adopted in the main text,
\begin{eqnarray}\label{sec1:df-error}
\expt{(\delta\theta)^2}={\textrm{Var}(\hat{\theta})},
\end{eqnarray}
for an unbiased estimator.

To estimate $\theta$ as precisely as possible,
one needs to optimize the estimation error $\expt{(\delta\theta)^2}$
over all estimators and measurements.
This is exactly the way that the QCRB was \GJ{proven},
which can be stated \GJ{via} the following two steps \cite{1994Braunstein3439,2009Paris125}.
The first step is to optimize $\expt{(\delta\theta)^2}$ over all estimators for a
given measurement. This leads to the CCRB,
\begin{eqnarray}\label{CCRB}
\expt{(\delta\theta)^2}\geq [{NF(\theta)}]^{-1},
\end{eqnarray}
where $F(\theta)$ is the CFI defined as
\begin{eqnarray}\label{CFI}
F(\theta)&=&\int dx\,p_\theta(x)\left(\frac{\partial}{\partial\theta}
\ln p_\theta(x)\right)^2\nonumber\\
&=&\int dx\,\frac{1}{p_\theta(x)}\left(\frac{\partial}{\partial\theta}
p_\theta(x)\right)^2.
\end{eqnarray}
The CCRB has been widely used in classical estimation theory. Its proof
can be found in many articles (e.g., Ref.~\cite{1994Braunstein3439})
and is definitely correct. We omit it here.
The second step is to optimize $F(\theta)$ over all measurements to
get the QCRB. This step is based on the following
equality \cite{1994Braunstein3439,2009Paris125},
\begin{eqnarray}\label{eq:part-p}
\frac{\partial}{\partial\theta} p_\theta(x)=\tr\left
(\Pi_x\frac{\partial}{\partial\theta}\rho_\theta\right),
\end{eqnarray}
which is valid if $\Pi_x$ satisfies the $\theta$-independence assumption,
as can be seen from Eq.~(\ref{df:p}).

To get the QCRB from Eq.~(\ref{eq:part-p}) (see also Ref.~\cite{2009Paris125}),
one needs to introduce the Symmetric Logarithmic Derivative (SLD), which is
defined as the Hermitian operator $L_\theta$ satisfying
\begin{eqnarray}\label{SLD}
\frac{L_\theta\rho_\theta+\rho_\theta L_\theta}{2}=
\frac{\partial\rho_\theta}{\partial\theta}.
\end{eqnarray}
Substituting Eq.~(\ref{SLD}) into Eq.~(\ref{eq:part-p}) yields
\begin{eqnarray}\label{sec1:s1}
\frac{\partial}{\partial\theta}p_\theta(x)
=\Re[\tr(\rho_\theta\Pi_x L_\theta)].
\end{eqnarray}
Inserting Eq.~(\ref{sec1:s1}) into Eq.~(\ref{CFI}) and
using Eq.~(\ref{df:p}), one has
\begin{eqnarray}
F(\theta)=\int dx\frac{\Re[\tr(\rho_\theta\Pi_x L_\theta)]^2}{\tr(\Pi_x\rho_\theta)},
\end{eqnarray}
which further leads to
\begin{eqnarray}\label{sec1:s2}
F(\theta)&\leq&\int dx\,\abs{\frac{\tr(\rho_\theta\Pi_x L_\theta)}
	{\sqrt{\tr(\Pi_x\rho_\theta)}}}^2\nonumber\\
&=&\int dx
\abs{\tr\left[\frac{\sqrt{\rho_\theta}\sqrt{\Pi_x}}{\sqrt{\tr(\Pi_x\rho_\theta)}}
	\sqrt{\Pi_x}L_\theta\sqrt{\rho_\theta}\right]}^2.
\end{eqnarray}
Using the Cauchy-Schwarz inequality
$\abs{\tr (X^\dagger Y)}^2\leq\tr(X^\dagger X)\tr(Y^\dagger Y)$ for
two matrices $X$ and $Y$, one can deduce from Eq.~(\ref{sec1:s2}) that
\begin{eqnarray}
F(\theta)\leq\int dx\,\tr(\Pi_xL_\theta\rho_\theta L_\theta)
=\tr(L_\theta\rho_\theta L_\theta)
=\tr(\rho_\theta L_\theta^2).\nonumber\\
\end{eqnarray}
So, the CFI is bounded from above by the so-called QFI
\begin{eqnarray}\label{CFI-QFI}
F(\theta)\leq H(\theta):=\tr(\rho_\theta L_\theta^2).
\end{eqnarray}
Substituting Eq.~(\ref{CFI-QFI}) into Eq.~(\ref{CCRB}), one arrives at the QCRB,
\begin{eqnarray}\label{QCRB}
\expt{(\delta\theta)^2}\geq [{NH(\theta)}]^{-1}.
\end{eqnarray}

It has never been doubted until recently \cite{2017Seveso12111} that the QCRB is the ultimate precision
allowed by quantum mechanics that cannot be surpassed under any circumstances \cite{1994Braunstein3439,2009Paris125}.
\textit{As can be seen from the above proof,
	Eqs.~(\ref{CFI-QFI}) and (\ref{QCRB}) as well
	as this belief are based on the $\theta$-independence assumption.}
At first glance, this assumption seems reasonable \GJ{as it echoes with  our quick
	impressions obtained from textbook materials
	regarding PMs}. Indeed, the textbook physics says that
the POVM operators associated with a PM
are the eigen-projections of the measured observable. An incautious follow-up
inference might be that the POVM operators associated with any types of measurements
are only determined by
the measured observable and, therefore, $\theta$-independent. However,
as critically examined below,
the POVM operators of a measurement may be determined by many factors in practice, even possibly
including the parameter $\theta$ itself. That is,
the $\theta$-independence assumption may not be true in practice. Therefore, although the CCRB is always true,
the QCRB as a universally valid bound
is questionable.

\section{Proof of optimality of Eq.~(\ref{Ex-error})}\label{app:B}

\begin{figure}[htbp]
	\includegraphics[width=0.4\textwidth]{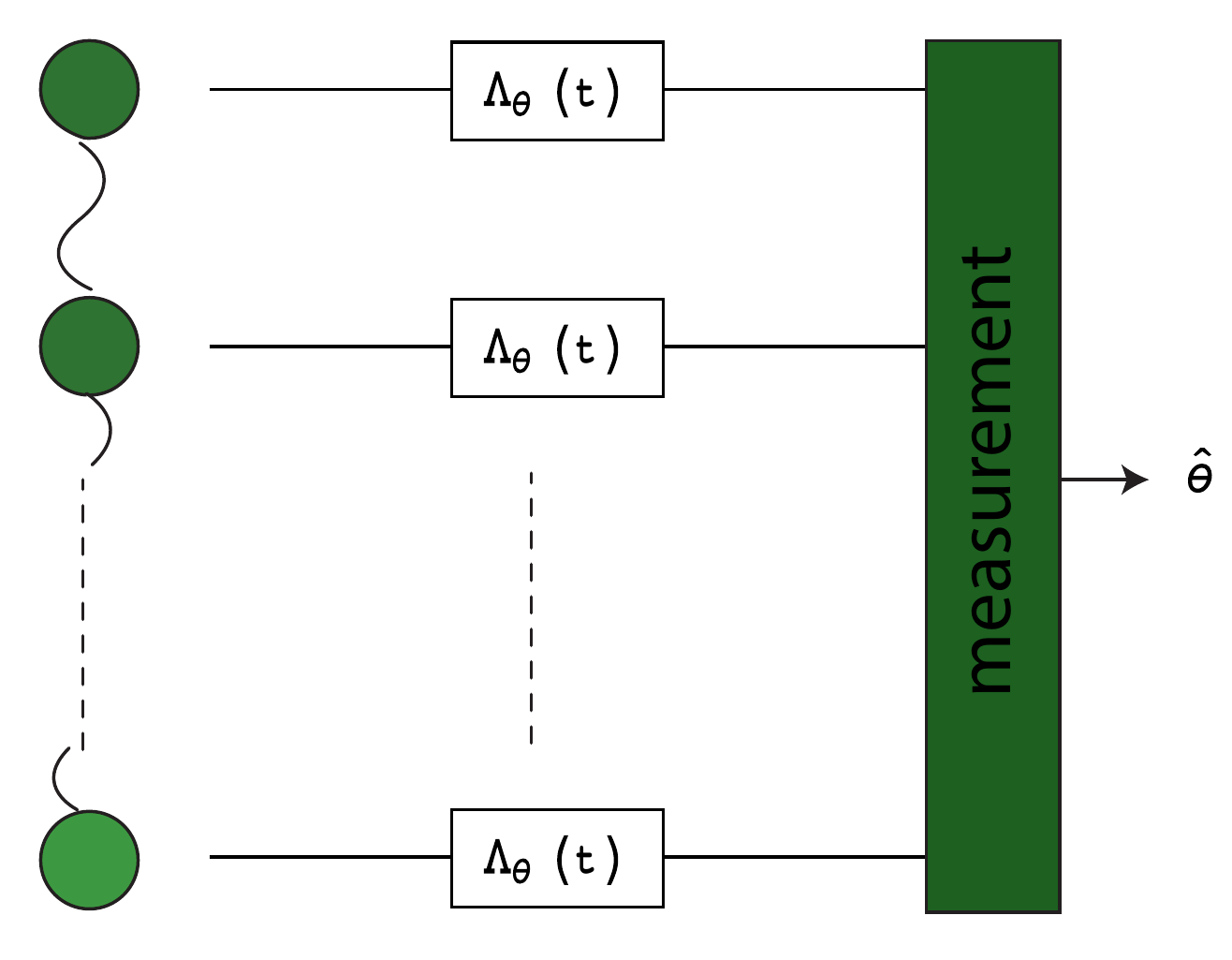}
	\caption{General scheme for quantum metrology.
		$N$ probes, prepared in an initial state, are sent through $N$ parallel
		channels $\Lambda_\theta(t)$. A measurement is performed on the final state,
		from which the parameter $\theta$ is estimated via an inference
		rule $\hat{\theta}$.}
	\label{sm-fig1}
\end{figure}

Here, we prove that Eq.~(\ref{Ex-error}) is optimal if the QCRB is valid.
Let
\begin{eqnarray}\label{sm-channel}
\Lambda_\theta(t):=e^{\mathcal{L}_\theta t}
\end{eqnarray}
be the quantum channel associated with the generalized amplitude damping process.
To estimate the parameter $\theta$ characterizing this channel,
a general strategy \cite{2006Giovannetti10401} is to send $N$ probes
through $N$ parallel channels $\Lambda_\theta(t)$, measure them at the output,
and use an inference rule $\hat{\theta}(\bm{x})$ to extract the value
of $\theta$ from the measurement results $\bm{x}$ (see Fig.~\ref{sm-fig1}).
Here, we have collectively denoted the measurement outcomes as $\bm{x}$, i.e.,
$\bm{x}=(x_1,\cdots,x_N)$.
Therefore, a general scheme of quantum metrology consists of
three ingredients: an input state of $N$ probes,
a measurement at the output, and an inference rule.
In the following, we do not impose any restriction on these ingredients.
That is, the input state can be an arbitrary, possibly highly entangled,
state; the measurement is a general, not necessarily local, POVM (which is,
of course, assumed to be $\theta$-independent so that the QCRB can be applied);
and the inference rule can be biased or unbiased.
Additionally, $t$ appearing in Eq.~(\ref{sm-channel}) can take any non-negative
value; it is unnecessary to fulfill the condition assumed in
the main text, which requires that the evolution time $t$ is much
longer than the relaxation time of the channel.

Using the QCRB \cite{1994Braunstein3439},
we have that the estimation error is lower bounded by
\begin{eqnarray}\label{Q-CR}
\expt{(\delta\theta)^2}\geq\frac{1}{{H\left[\Lambda_\theta^{\otimes N}(t)[\rho(0)]\right]}}.
\end{eqnarray}
Here, $H\left[\Lambda_\theta^{\otimes N}(t)[\rho(0)]\right]$ represents the QFI
and $\Lambda_\theta^{\otimes N}(t)[\rho(0)]$ is the output state of the $N$ probes,
where $\rho(0)$ denotes the input state.
To evaluate the QFI $H\left[\Lambda_\theta^{\otimes N}(t)[\rho(0)]\right]$, we
introduce two amplitude damping channels
\cite{2010Nielsen}, $\Lambda_i(t)$, $i=0,1$, defined as two completely positive
trace-preserving (CPTP) maps transforming the Bloch vector as
\begin{eqnarray}
&&\Lambda_0(t):\ \ (r_x,r_y,r_z)\rightarrow\nonumber\\
&&(e^{-\frac{\gamma t}{2}}r_x,
e^{-\frac{\gamma t}{2}}r_y,1-e^{-\gamma t}+e^{-\gamma t}r_z),
\end{eqnarray}
and
\begin{eqnarray}
&&\Lambda_1(t):\ \ (r_x,r_y,r_z)\rightarrow\nonumber\\
&&(e^{-\frac{\gamma t}{2}}r_x,
e^{-\frac{\gamma t}{2}}r_y,e^{-\gamma t}-1+e^{-\gamma t}r_z),
\end{eqnarray}
respectively.
Noting that the effect of $\Lambda_\theta(t)$ is
\begin{eqnarray}
&&\Lambda_\theta(t):\ \ (r_x,r_y,r_z)\rightarrow\nonumber\\
&&\left(e^{-\frac{\gamma t}{2}}r_x,
e^{-\frac{\gamma t}{2}}r_y,(2\theta-1)(1-e^{-\gamma t})+e^{-\gamma t}r_z\right),
\nonumber\\
\end{eqnarray}
we have
\begin{eqnarray}\label{eq:decom}
\Lambda_\theta(t)=\theta\Lambda_0(t)+(1-\theta)\Lambda_1(t).
\end{eqnarray}
Equation (\ref{eq:decom}) enables us to rewrite $\Lambda_\theta(t)$ as
a $\theta$-independent quantum channel acting on a larger input space,
\begin{eqnarray}\label{eq:big-map}
\Lambda_\theta(t)[\rho]=\Phi(t)[\rho\otimes\rho_\theta].
\end{eqnarray}
Here, $\rho_\theta=\textrm{diag}(\theta,1-\theta)$ is the steady state,
and $\Phi(t)$ is defined as
\begin{eqnarray}
\Phi(t)[\rho\otimes\sigma]&:=&\sum_{i=0,1}\Lambda_i(t)\otimes\mathcal{E}_i
[\rho\otimes\sigma]\nonumber\\
&=&\sum_{i=0,1}\Lambda_i(t)[\rho]\otimes\mathcal{E}_i
[\sigma],
\end{eqnarray}
where $\mathcal{E}_i[\sigma]:=\bra{i}\sigma\ket{i}$, $i=0,1$.
It is not difficult to see that $\Phi(t)$ thus
defined is
a CPTP map.
Using Eq.~(\ref{eq:big-map}), we have
\begin{eqnarray}
H\left[\Lambda_\theta^{\otimes N}(t)[\rho(0)]\right]&=&
H\left[\Phi^{\otimes N}(t)[\rho(0)\otimes\rho_\theta^{\otimes N}]\right]\nonumber\\
&\leq& H\left[\rho(0)\otimes\rho_\theta^{\otimes N}\right]\nonumber\\
&=&H\left[\rho_\theta^{\otimes N}\right],
\end{eqnarray}
where we have used the monotonicity of the QFI under parameter-independent
CPTP maps \cite{1994Braunstein3439}.
Noting that $H\left[\rho_\theta^{\otimes N}\right]=NH\left[\rho_\theta\right]$
and $H[\rho_\theta]=\frac{1}{\theta(1-\theta)}$, we further have
\begin{eqnarray}\label{eq:bound-F}
H\left[\Lambda_\theta^{\otimes N}(t)[\rho(0)]\right]\leq\frac{N}{\theta(1-\theta)}.
\end{eqnarray}
Substituting Eq.~(\ref{eq:bound-F}) into Eq.~(\ref{Q-CR}) yields
\begin{eqnarray}
\expt{(\delta\theta)^2}\geq{\frac{\theta(1-\theta)}{N}},
\end{eqnarray}
indicating that Eq.~(\ref{Ex-error}) is optimal if the QCRB is valid.

\section{More details on the minimal model of quantum measurement}\label{app:C}

Suppose that $\mathscr{S}$
is a qubit undergoing the generalized amplitude damping process,
\begin{eqnarray}\label{sec2:model}
\frac{d}{d t}\rho_\mathscr{S}(t)=\mathcal{L}_\theta\rho_\mathscr{S}(t),
\end{eqnarray}
with $\mathcal{L}_\theta$ defined in the main text.
Equation (\ref{sec2:model}) can
be used to describe the dynamics of a qubit interacting with a
Bosonic thermal environment at finite temperature \cite{2010Nielsen}.
In this scenario, $\theta$ is a monotone function of the temperature of
the environment, which characterizes losses of energy from the qubit,
i.e., effects of energy dissipation \cite{2010Nielsen}.
So, determining $\theta$ amounts to determining the temperature
of the environment. Considering that temperature estimation has received
much attention recently in quantum thermometry \cite{2017Mancino130502,2019Mehboudi30403,2019Seah180602},
we aim to estimate $\theta$ in the main text. We assume that the
dynamics described by Eq.~(\ref{sec2:model}) is always-on.
Such an assumption is, of course, reasonable in many physical
scenarios/applications, e.g., in quantum computation,
where effects of decoherence and dissipation are always-on whenever
one implements unitary gates or performs quantum measurements on qubits.

To simplify our discussion here as well as in the main text,
we adopt the minimal model of quantum measurement, which has been used
time and again in the literature \cite{2019Naghiloo}. That is, to measure an observable $A$,
we add an interaction term,
\begin{eqnarray}\label{interaction}
H_I=T^{-1}A\otimes\hat{p},
\end{eqnarray}
coupling $\mathscr{S}$ to a measuring apparatus $\mathscr{A}$.
Here and in the main text, we have suppressed the subscript $x$ in $\hat{p}$ for ease of notation.
The initial state $\ket{\phi}$ of $\mathscr{A}$ is set to be a Gaussian centered
at $x = 0 $,
\begin{eqnarray}
\ket{\phi}=\int dx \frac{1}{(2\pi\sigma^2)^{1/4}}e^{-\frac{x^2}{4\sigma^2}}\ket{x},
\end{eqnarray}
where $\sigma$ denotes the standard deviation.
Here and henceforth, we omit the two limits of a integral
whenever they are $-\infty$ and $+\infty$ for ease of notation.
As usual, we are not interested in the free dynamics of $\mathscr{A}$ itself. So, we require the free Hamiltonian of $\mathscr{A}$
to be zero. Such a requirement is often imposed in proposals of
quantum measurements and may be satisfied if one goes to a frame that
rotates with the free Hamiltonian of $\mathscr{A}$ in the rest frame
(see p.~15 of Ref.~\cite{2019Naghiloo}). It is worth noting that there are a
number of experimental setups (such as in cavity quantum electrodynamics and
circuit quantum electrodynamics) that are physically \GJ{equivalent} and therefore
can be used to implement the above minimal model  \cite{2019Naghiloo}. In particular,
the above minimal model was also adopted in Aharonov \textit{et al.}'s proposal of
adiabatic measurements \cite{1993Aharonov38}, which has been experimentally
realized in optical setups \cite{2017Piacentini1191}.

\begin{figure}[htbp]
	\includegraphics[width=0.4\textwidth]{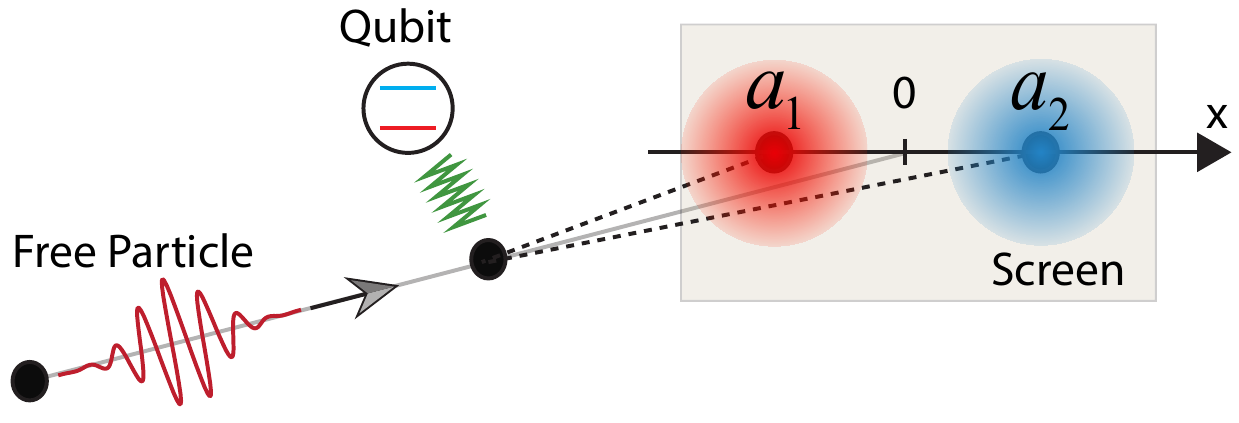}
	\caption{A toy setup used to demonstrate the implementation of PM (see p.~73 of Ref.~\cite{2019Naghiloo}). A free particle passes by and interacts with a
		qubit. The state of the free particle is described by a Gaussian wave packet moving along $z$ direction. Upon interacting with the qubit, the particle is pulled or pushed along $x$ direction, depending on the state of the qubit. In a PM, the position of the
		particle hitting the screen tells us about the eigenvalues of the measured observable $A$. However, in a DAM, the shift is the expectation value $\expt{A}_\theta$.}
	\label{fig-sm-toy-model}
\end{figure}

It may be helpful to recall a toy setup used to demonstrate the implementation of PM \cite{2019Naghiloo}. As schematically shown in Fig.~\ref{fig-sm-toy-model},
a free particle passing by the qubit plays the role of the measuring apparatus $\mathscr{A}$. Initially, it is prepared in a Gaussian wave packet with a non-zero momentum along $z$ direction,
\begin{eqnarray}
\phi(\bm{r},0)=\frac{1}{(2\pi\sigma^2)^{3/4}}e^{-\frac{x^2+y^2+z^2}
	{4\sigma^2}} e^{ip_z z},
\end{eqnarray}
where $p_z$ is a fixed positive number describing the $z$-component of the momentum of the particle. Then the wave packet moves along $z$ direction with the group velocity
\begin{eqnarray}
v_g=\frac{p_z}{m},
\end{eqnarray}
where $m$ denotes the mass of the particle. That is,
\begin{eqnarray}
\phi(\bm{r},t)=\frac{1}{(2\pi\sigma^2)^{3/4}}e^{-\frac{x^2+y^2+(z-v_g t)^2}
	{4\sigma^2}} e^{i(p_z z-\omega t)},
\end{eqnarray}
with $\omega=p_z^2/(2m)$ ($\hbar=1$). Here, we have neglected the spread of the wave packet by assuming that $m$ is very large. Note that $v_g$ can take an arbitrary value because of the freedom in choosing $p_z$.
Upon interacting with the qubit, the particle gets pulled or pushed along $x$ direction, depending on the state of the qubit. That is, the interaction is of the form $H_I=g (t)A\otimes \hat{p}_x$. [In Eq.~(\ref{interaction}), $g(t)$ is assumed to be time-independent for simplicity.] Taking $A=\sigma_z$ as an example, we see that whether the particle gets pulled or pushed depends on whether the state of the qubit is $\ket{0}$ or $\ket{1}$. The coupling strength is determined by the distance of the particle from the qubit, whereas the coupling time is determined by the value of $v_g$ as well as the distance between the particle and the screen. In a PM, the coupling strength is large and the coupling time is small, so that $H_I$ is dominant whereas  $\mathcal{L}_\theta$ can be neglected. Under the influence of $H_I$, the particle hits the screen with some shift along $x$ direction. The shift tells us about an eigenvalue $a_i$ of $A$. On the other hand, in a DAM, we consider the scenario that $\mathcal{L}_\theta$ is dominant but $H_I$ is comparatively weak. The former effectively reshapes the latter to be $g(t)\expt{A}_\theta\hat{p}_x$. So, the shift is now the expectation value $\expt{A}_\theta$ rather than an eigenvalue $a_i$.

According to the minimal model of measurement and noting that the dynamics (\ref{sec2:model}) is always-on, we have that the dynamical equation describing
the coupling of $\mathscr{S}$ and $\mathscr{A}$ reads
\begin{eqnarray}\label{sm-eq:measuring}
\frac{d}{dt}\rho_{\mathscr{SA}}(t)=\mathcal{L}_\theta\rho_{\mathscr{SA}}(t)-i
\left[H_I,\rho_{\mathscr{SA}}(t)\right]
=:\mathcal{L}\rho_{\mathscr{SA}}(t).\nonumber\\
\end{eqnarray}
Evidently, the dynamics map associated with the coupling procedure is
\begin{eqnarray}
\mathcal{E}_T=e^{\mathcal{L}T},
\end{eqnarray}
transforming the initial state $\rho_\theta\otimes\ket{\phi}\bra{\phi}$ to
the state $\mathcal{E}_T(\rho_\theta\otimes\ket{\phi}\bra{\phi})$ at time $T$. In the main text, we mainly focus on the two limits $T\rightarrow 0$ and $T\rightarrow\infty$, which correspond to PMs and DAMs, respectively. Besides, although $T$ is set to be $\infty$ in DAMs for the sake of mathematical convenience, it is not very large for strongly dissipative systems. Moreover, the larger the dissipative gap is, the shorter $T$ can be. This is analogous to the well known result regarding the adiabatic theorem, where $T$ is determined by the energy gap of the Hamiltonian of a closed system.

\section{Discussion on intermediate $T$} \label{app:D}
\begin{figure}[H]
	\centering
	\subfigure[]{
		\includegraphics[width=.22\textwidth]{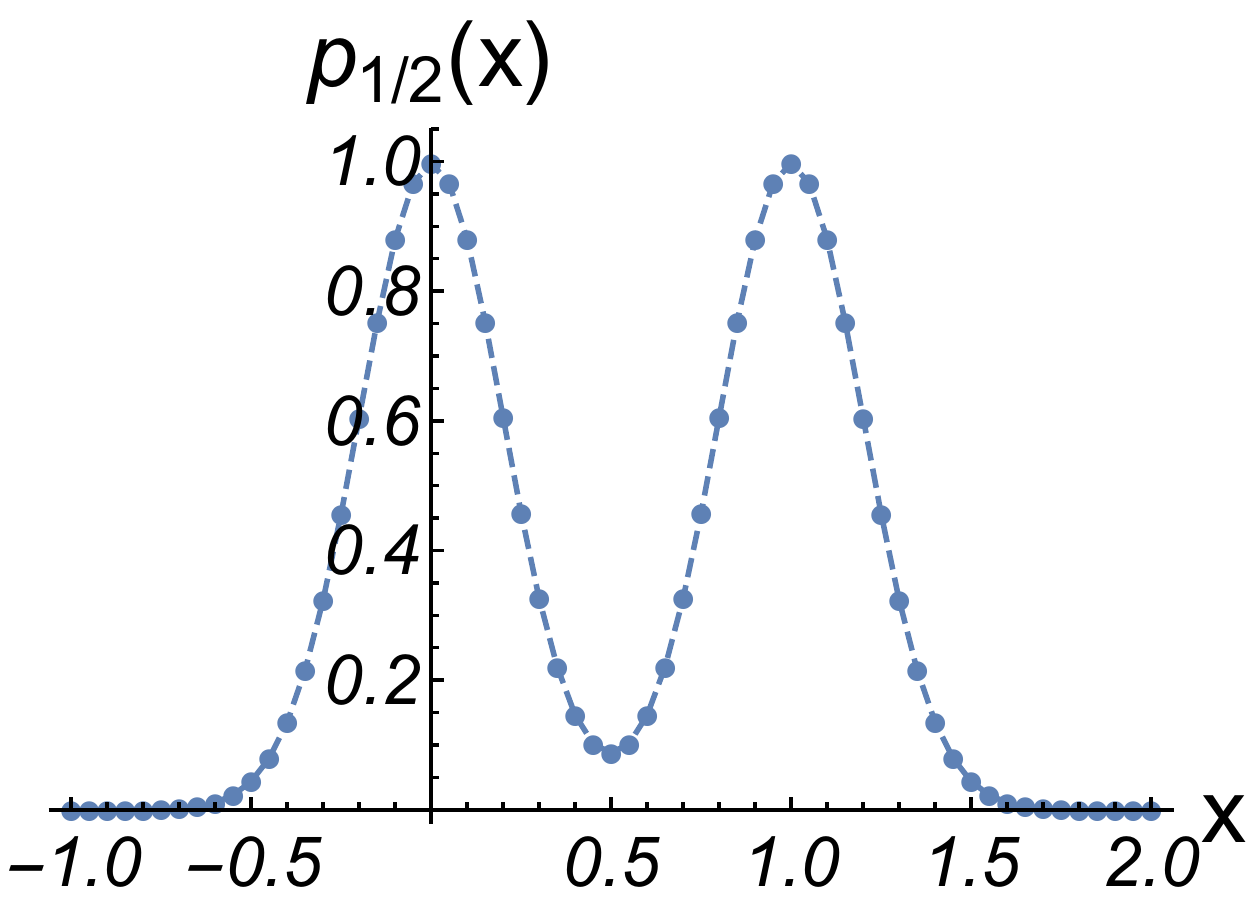}\label{fig-generic-1}}
	\subfigure[]{
		\includegraphics[width=.22\textwidth]{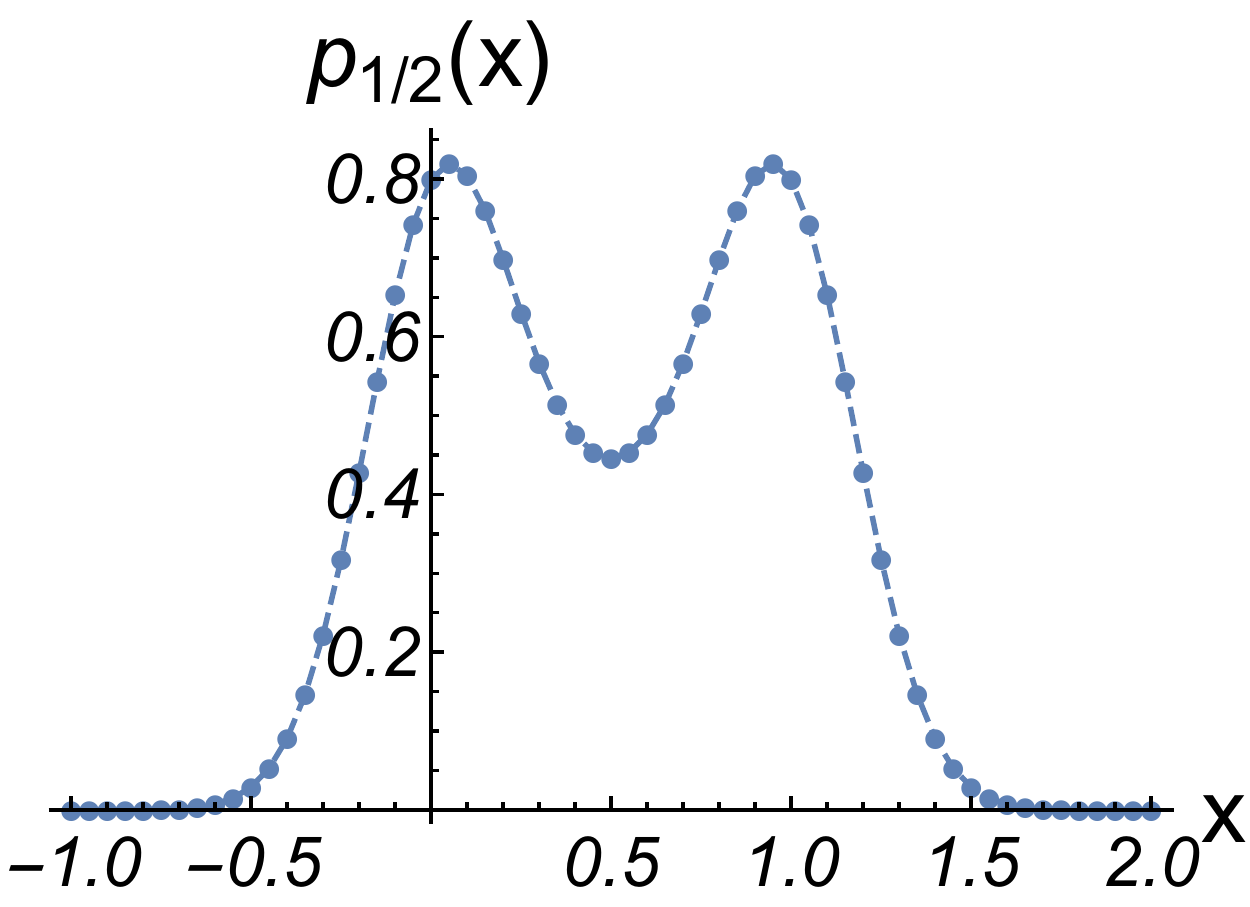}\label{fig-generic-2}}\\
	\subfigure[]{
		\includegraphics[width=.22\textwidth]{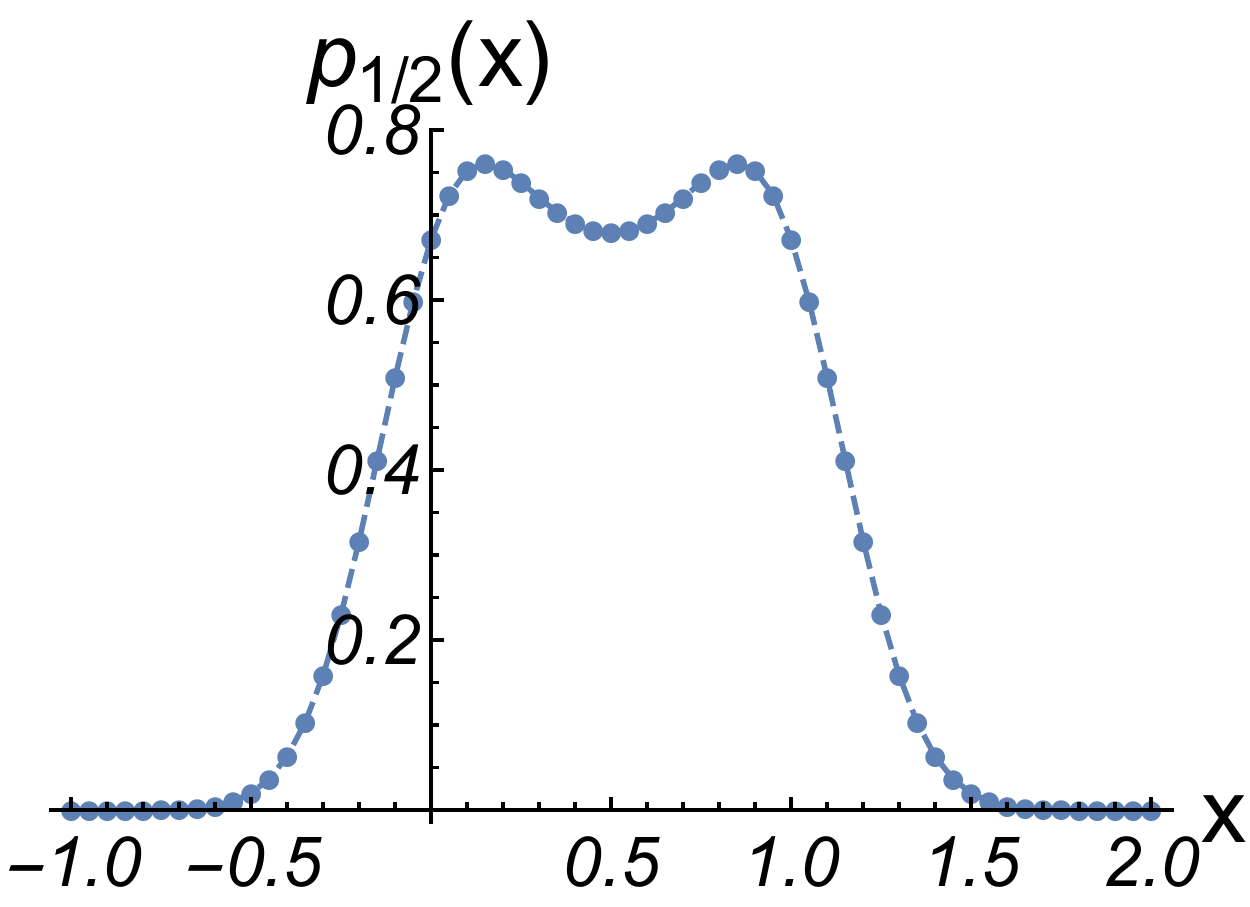}\label{fig-generic-3}}
	\subfigure[]{
		\includegraphics[width=.22\textwidth]{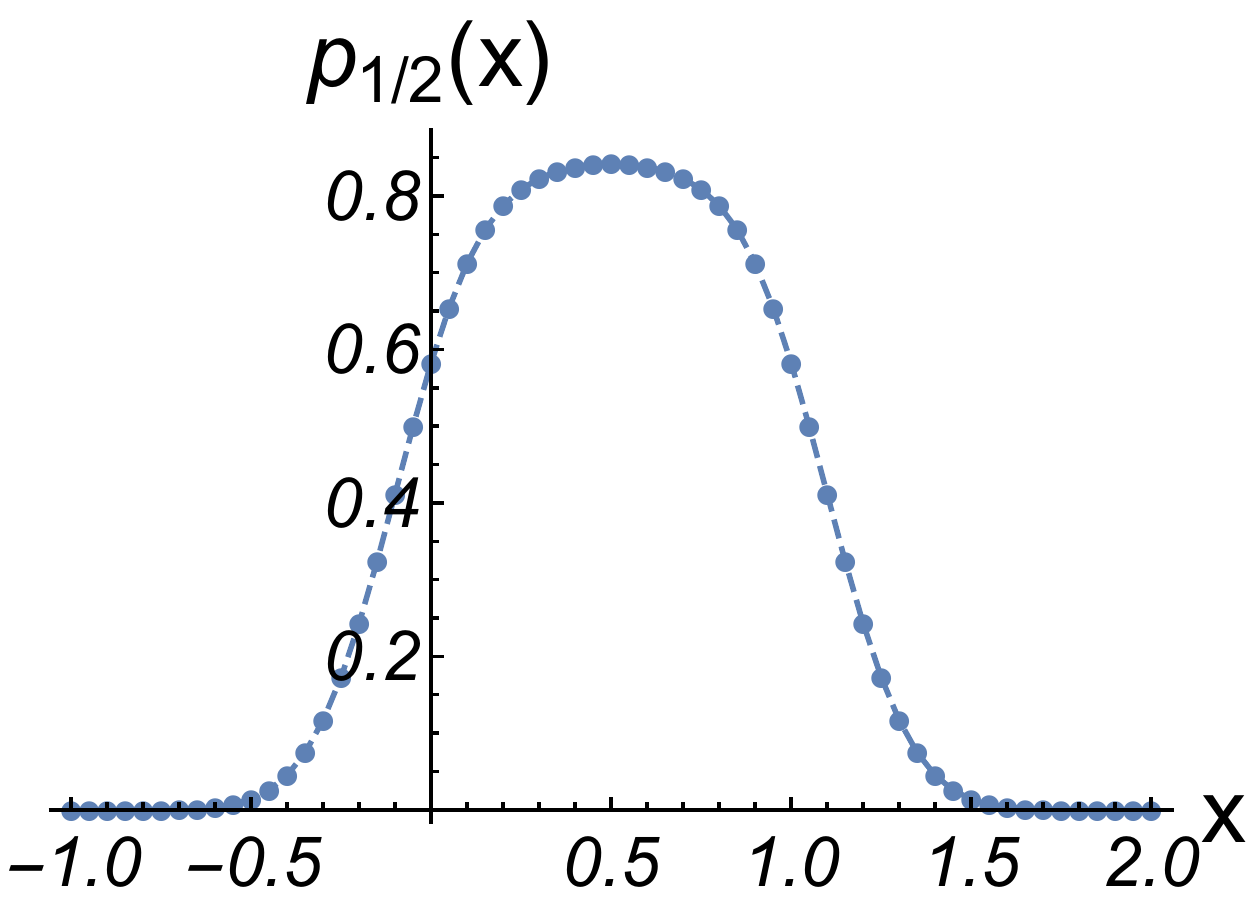}\label{fig-generic-4}}\\
	\subfigure[]{
		\includegraphics[width=.22\textwidth]{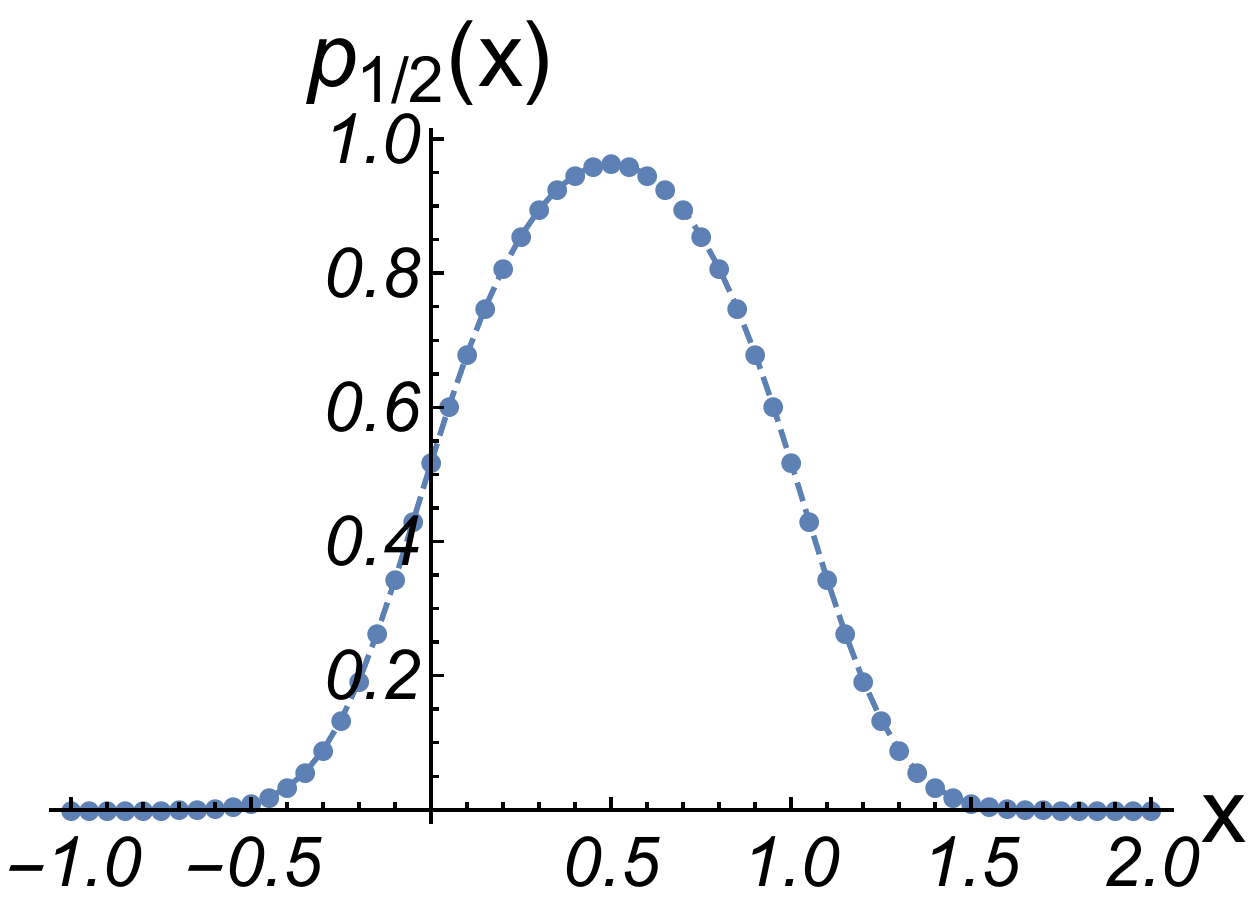}\label{fig-generic-5}}
	\subfigure[]{
		\includegraphics[width=.22\textwidth]{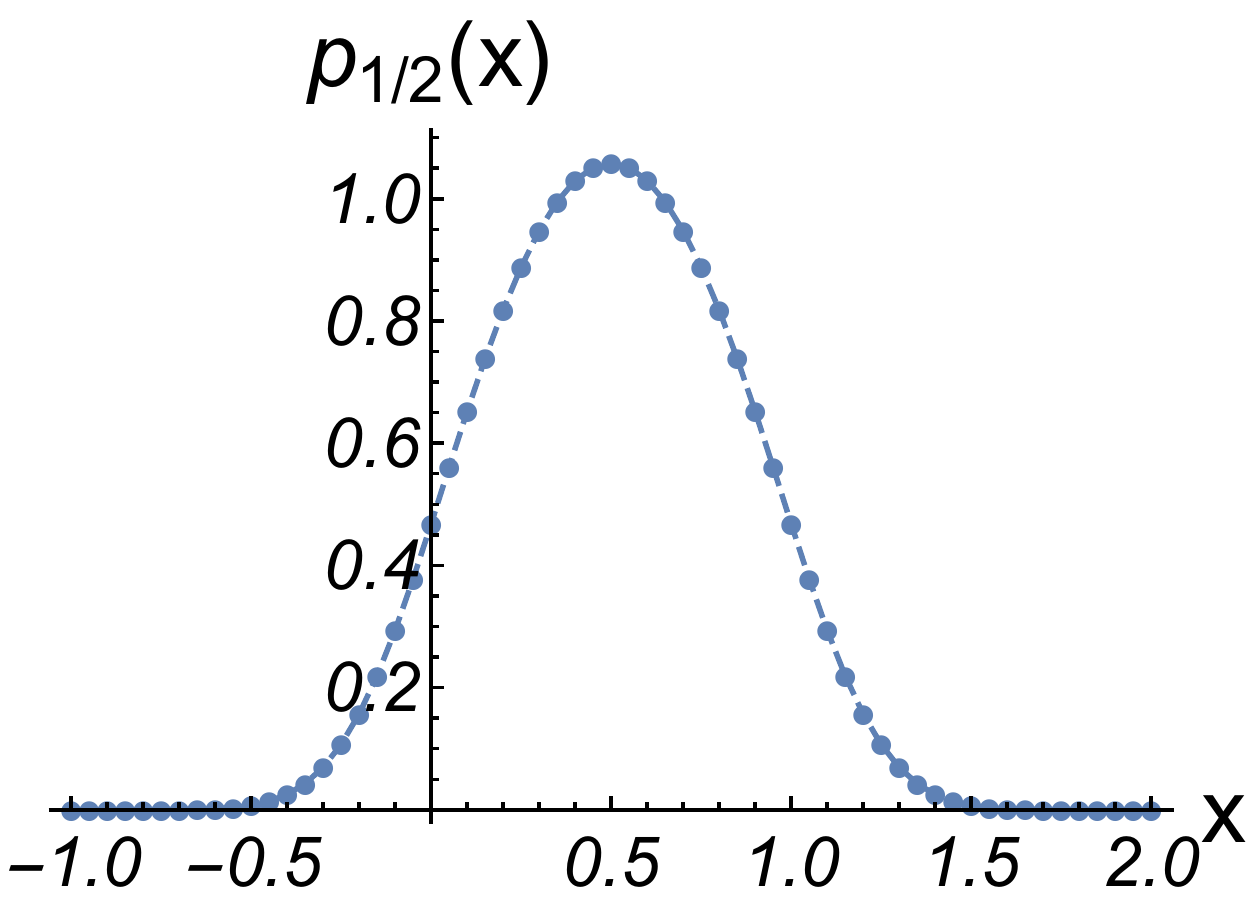}\label{fig-generic-6}}\\
	\subfigure[]{
		\includegraphics[width=.22\textwidth]{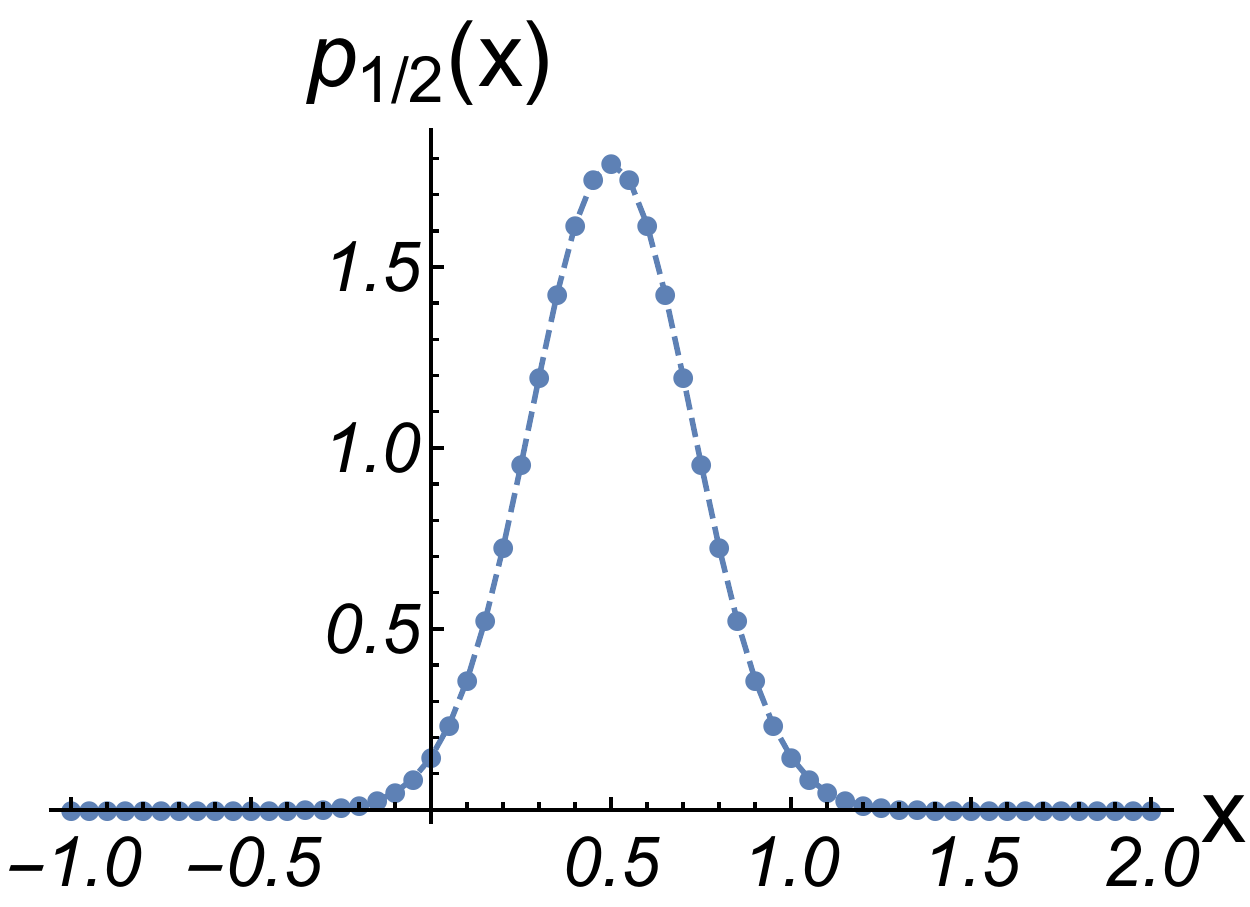}\label{fig-generic-7}}
	\subfigure[]{
		\includegraphics[width=.22\textwidth]{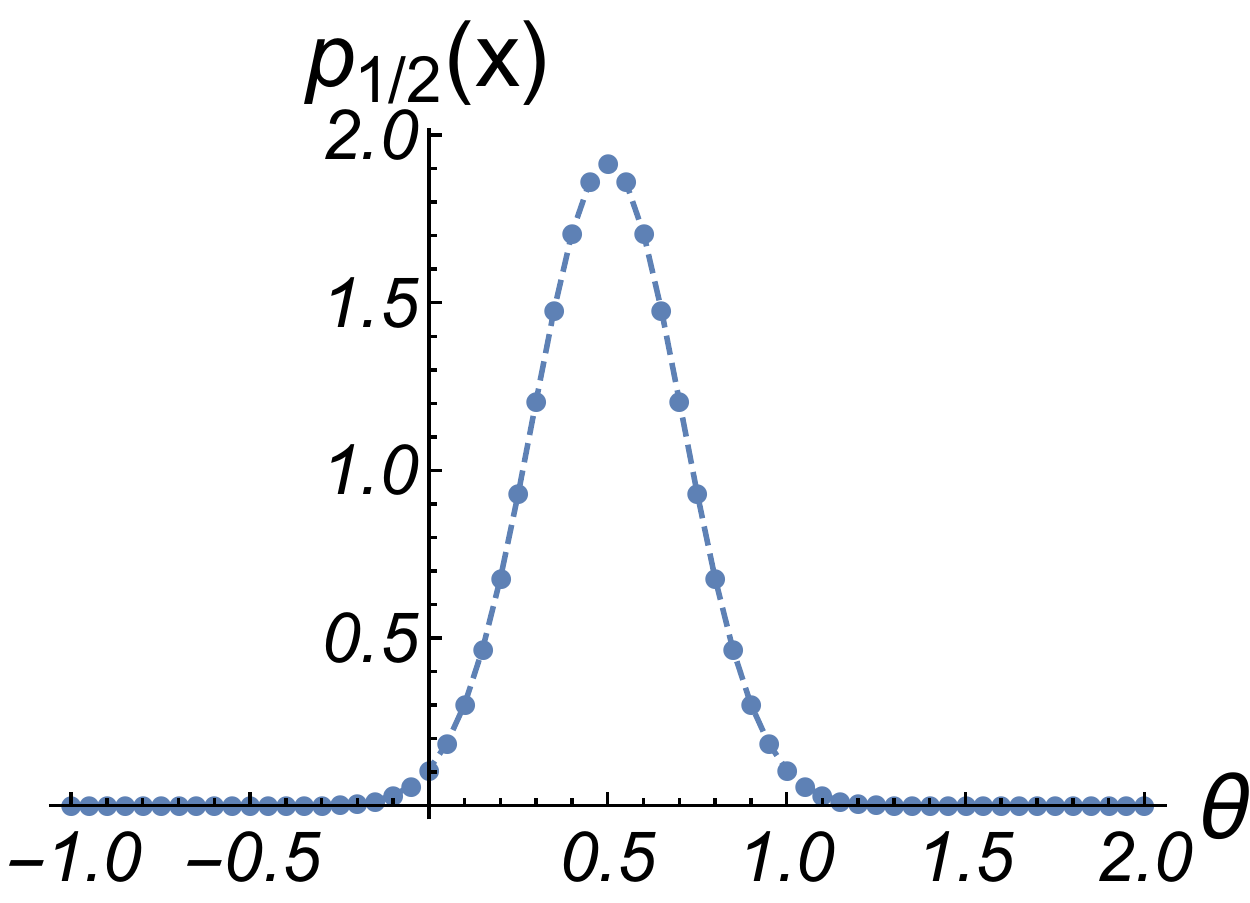}\label{fig-generic-8}}\\
	\subfigure[]{
		\includegraphics[width=.22\textwidth]{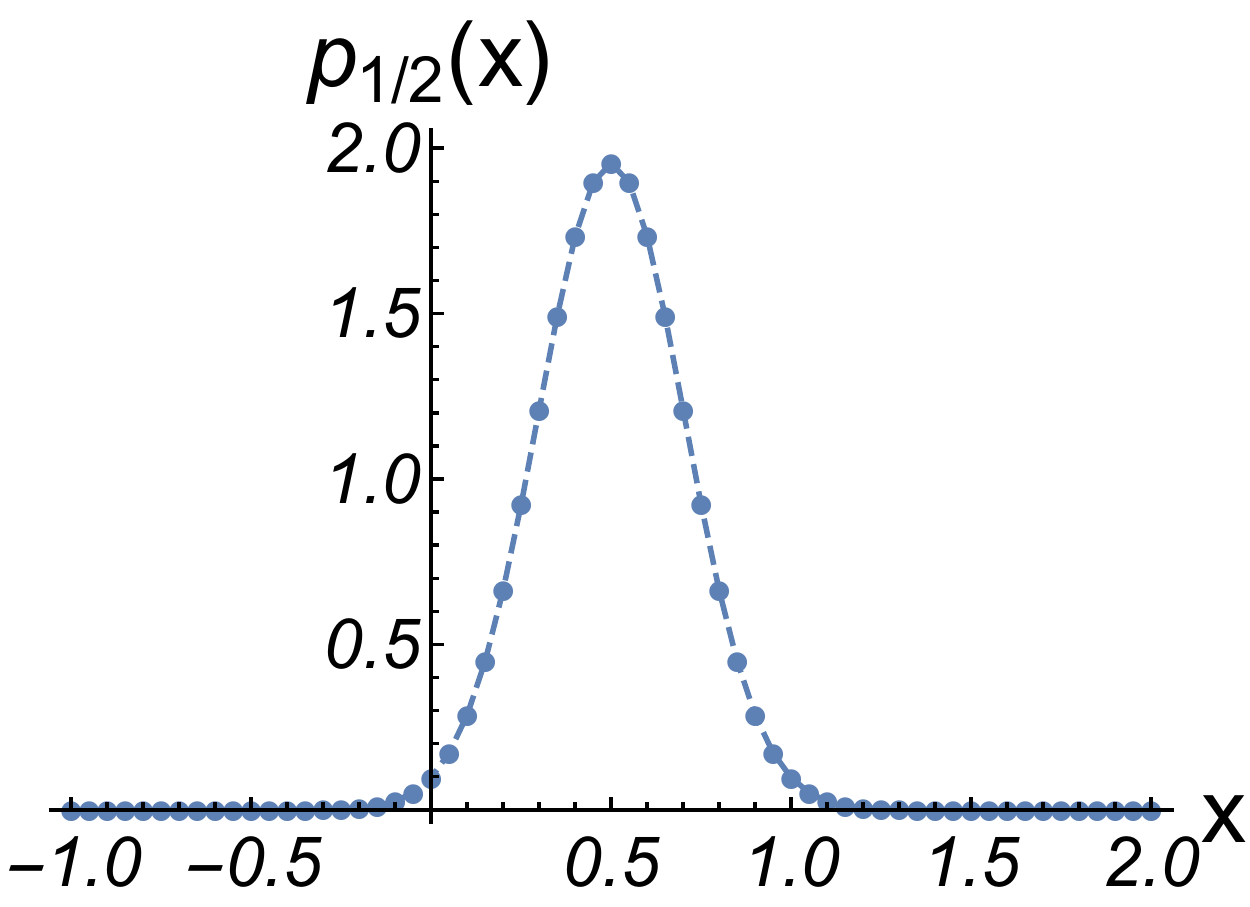}\label{fig-generic-9}}
	\subfigure[]{
		\includegraphics[width=.22\textwidth]{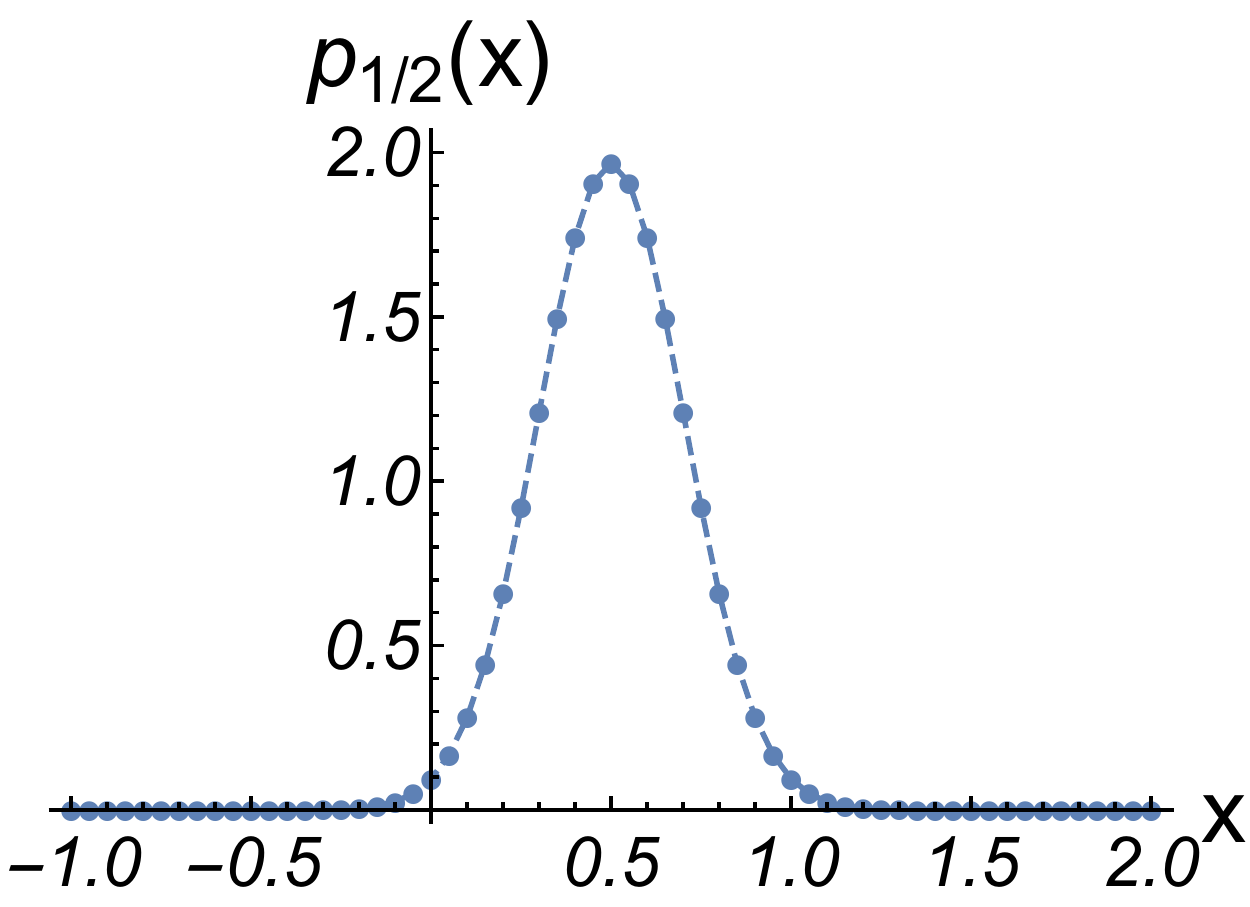}\label{fig-generic-10}}
	\caption{Plots of the probability distribution $p_{1/2}(x)$ \textcolor{black}{for (a) $T=0$ $\mu$s, (b) $T=0.2$ $\mu$s, (c) $T=0.4$ $\mu$s, (d) $T=0.6$ $\mu$s, (e) $T=0.8$ $\mu$s, (f) $T=1$ $\mu$s, (g) $T=10$ $\mu$s, (h) $T=30$ $\mu$s, (i) $T=60$ $\mu$s, and (j) $T=90$ $\mu$s}.
		Parameters used are $\sigma=0.2$ and $\gamma=5$ MHz.}
	\label{fig-generic}
\end{figure}
In the main text, we discussed the two limiting cases of $T\rightarrow 0$
and $T\rightarrow\infty$, which correspond to PM and
DAM, respectively. 
The former is
the best known type of quantum measurement, which has been extensively studied in
the literature. The physical significance of
the latter, however, has not
been widely appreciated or even fully recognized. The aim of the present work is
to study the limiting case of $T\rightarrow \infty$ and show that it is also
a novel type of quantum measurement \GJ{applied to dissipative systems}.
Here, we would like to go \GJ{one} step further and
briefly discuss the case of intermediate $T$,
although such a discussion is beyond the focus of this work. To do this,
we numerically
compute the profile of the probability distribution $p_\theta(x)$ for various values
of $T$.
Some typical examples of the profile are shown in Fig.~\ref{fig-generic}.
Here, we choose $\theta=1/2$ without loss of generality.
The profile associated with the PM is shown in
Fig.~\ref{fig-generic-1}, in which there are two peaks
corresponding to the two eigenvalues $0$ and $1$ of the measured observable $A_\textrm{opt}$.
As $T$ increases, the probability density at $x=\expt{A_\textrm{opt}}_\theta=1/2$
increases whereas those at $x=0$ and $x=1$ decrease,
as shown in Figs.~\ref{fig-generic-1}-\ref{fig-generic-4}.
This leads to the fact that the two peaks in Fig.~\ref{fig-generic-1} finally merge
into one peak, as shown in Fig.~\ref{fig-generic-4}. After that,
the probability density at $x=\expt{A_\textrm{opt}}_\theta=1/2$ continues to increase as $T$ increases,
so that the peak becomes sharper and sharper, as can be seen from
Figs.~\ref{fig-generic-4}-\ref{fig-generic-8}. Finally, when $T\geq 30~ \mu s$,
the profile rarely changes even though
we continue to enlarge $T$, as can be seen from Figs.~\ref{fig-generic-8}-\ref{fig-generic-10}.
This indicates that $T=30~ \mu s$
can be already thought of as being sufficiently large so that
the profile now corresponds to the DAM.

\section{Discussion on a realistic model}\label{app:E}

Here we examine the realistic model of a driven superconducting qubit dispersively coupled to a microwave
resonator \cite{2004Blais62320,2004Wallraff162}. It is equivalent to the adopted
minimal model of measurement and has been used to realize PMs \cite{2013Hatridge178}.
In a frame rotating at the frequency of the driving pulse, the Hamiltonian of the
qubit reads $H_q=-\frac{\omega_{R0}}{2}(\cos\alpha\sigma_x+\sin\alpha\sigma_y)-
\frac{\Delta
	\omega}{2}\sigma_z$, where $\sigma_i$, $i=x,y,z$, denote Pauli matrices,
and $\omega_{R0}$, $\Delta\omega$, and $\alpha$ are the Rabi frequency,
the detuning, and the phase of the driving pulse, respectively \cite{2005Ithier134519}.
The qubit is exposed to decoherence and dissipation described by
$\mathcal{L}_d=\gamma_1
\mathcal{D}[\sigma_-]+\frac{\gamma_2}{2}\mathcal{D}[\sigma_z]$ with
$\mathcal{D}[o]\rho=o\rho o^\dagger-\frac{1}{2}\{o^\dagger o,\rho\}$,
where $\gamma_i$, $i=1,2$, are the relaxation and
dephasing rates, respectively. The total Liouvillian describing the dynamics of the qubit therefore is $\mathcal{L}_\textrm{total}:=-i[H_q,\bullet]+\mathcal{L}_d$, which plays the role of $\mathcal{L}_\theta$ in Eq.~(\ref{sm-eq:measuring}).
The interaction between the qubit and the resonator reads $H_I=\chi a^\dagger a\sigma_z$,
where $\chi$ represents the dispersive coupling, and $a^\dagger$ and $a$ are the creation and annihilation operators for the resonator.
The Hamiltonian of the resonator is
$H_r=\omega_r a^\dagger a$, where $\omega_r$ is the resonator frequency. In PMs, under the influence of $H_I$, the frequency of the resonator is shifted as $\omega_r\pm\chi$, depending on whether the state of the qubit is $\ket{0}$ or $\ket{1}$. (In DAMs, it should be $\omega_r+\chi\expt{A}$.) The shift can be read out by coupling the resonator to transmission lines. The average number $\overline{n}$ of photons in the resonator is determined by the power of the readout pulse as well as the decoherence and dissipation experienced by the resonator. Here, we consider the scenario that the decoherence and dissipation of the qubit is much stronger than those of the resonator so that we can neglect the latter within a certain time window. On the other hand, we can suppress the term $H_r$ by switching to a rotating frame, that is, we are in a doubly rotating frame. So, we may only consider the two terms $\mathcal{L}_\textrm{total}$ and $H_I$ as in Eq.~(\ref{sm-eq:measuring}), in order to figure out the frequency shift resulting from the non-trivial interplay between them.
A full analysis taking into account the decoherence and dissipation of the resonator (possibly with experimental demonstration) is left to a future work. 
\begin{figure}
	\centering
	\vspace{0.2cm}
	\subfigure[]{
		\includegraphics[width=.22\textwidth]{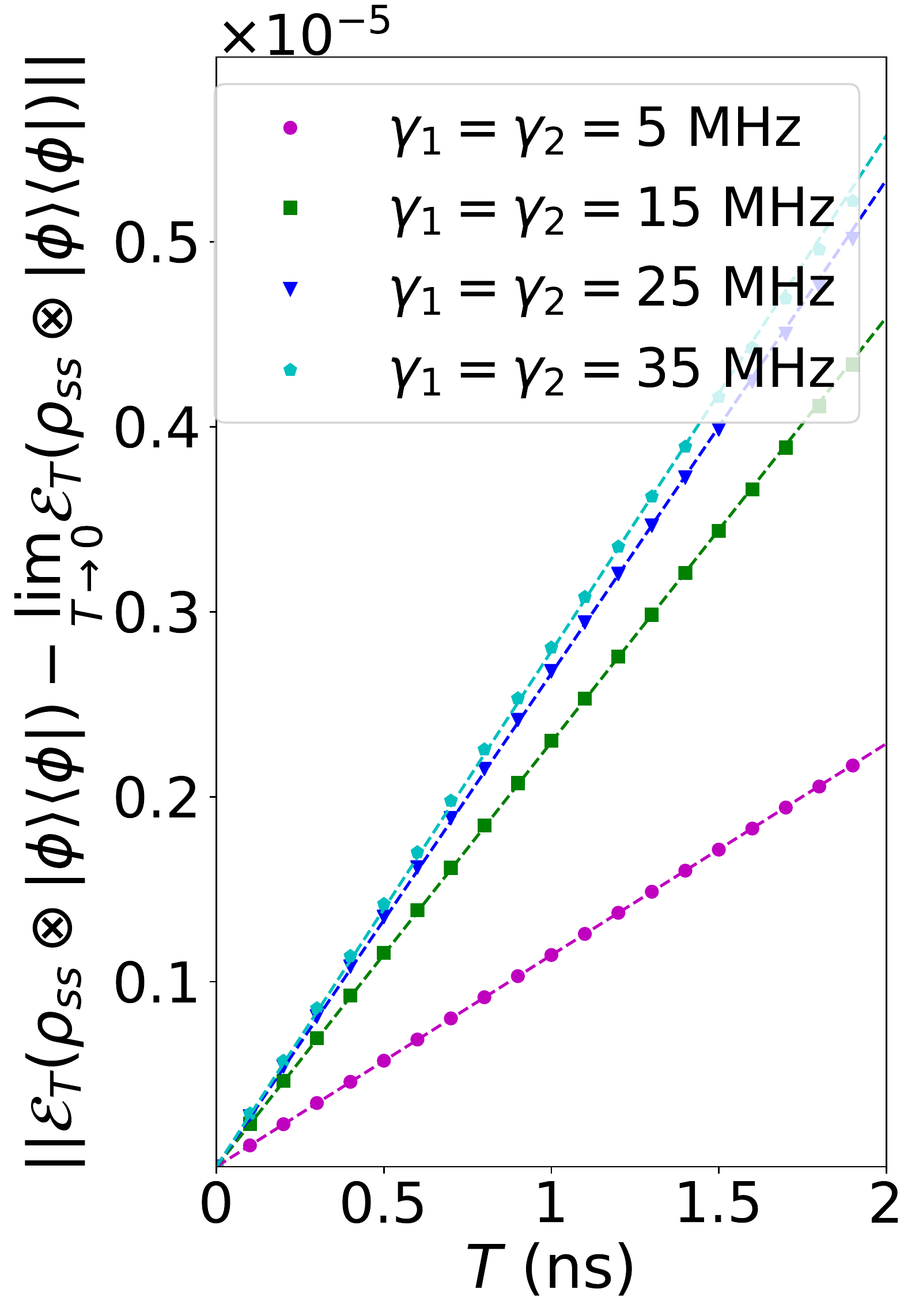}\label{fig2a-sm}}
	\subfigure[]{
		\includegraphics[width=.22\textwidth]{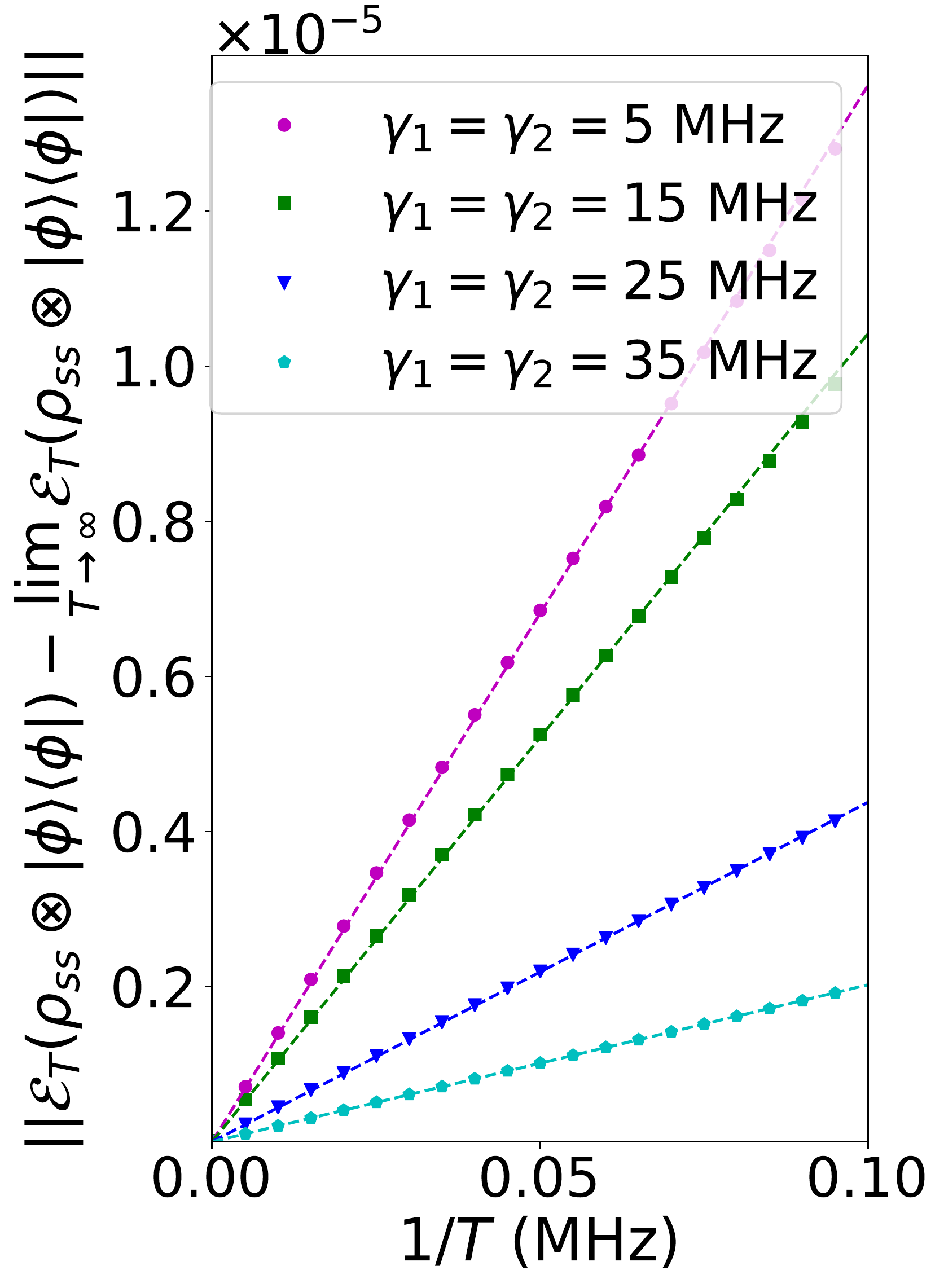}\label{fig2b-sm}}
	\caption{Numerical results of measures (\ref{measure-PM}) and (\ref{measure-DAM}) \textcolor{black}{for (a) PM and (b) DAM} with different $\gamma_1$ and $\gamma_2$. Parameters used are: $\omega_{R0}/(2\pi)=2$ MHz, $\alpha=0$, $\Delta\omega=0$, and $\overline{n}=16$.}
	\label{fig2-sm}
\end{figure}
Figure \ref{fig2-sm} shows the numerical results of the measures in Eqs.~(\ref{measure-PM}) and (\ref{measure-DAM}) for different $\gamma_1$ and $\gamma_2$. Here, we set $\gamma_1=\gamma_2$ without loss of generality. Also, we identify $\chi$ with $T^{-1}$, for the sake of notational consistency. As can be seen from Fig.~\ref{fig2-sm}, as $\gamma_1=\gamma_2$ increases, the slope of measure (\ref{measure-PM}) as a function of $T$ increases, but that of measure (\ref{measure-DAM}) as a function of $1/T$ decreases. This indicates that the strong interaction requirement in PMs is increasingly difficult to meet if decoherence and dissipation become increasingly strong, whereas this is not the case for the long time requirement in DAMs. For instance, if the desired tolerance of the deviations is set to be $10^{-5}$, $1/T$ in PM are 114, 230, 267, 279 MHz whereas $T$ in DAM are 13.6, 10.4, 4.4, 2.0 $\mu$s, for
$\gamma_1=\gamma_2=5,15,25,35$ MHz, respectively.
This is consistent with the results found in the main text.

\section{Details on examination of the $\theta$-independence assumption}\label{app:F}

To examine the $\theta$-independence assumption, we need to figure
out an expression for the POVM operator $\Pi_x$. Noting that the state
of $\mathscr{S}+\mathscr{A}$ immediately after the coupling procedure
is $\mathcal{E}_T(\rho_\theta
\otimes
\ket{\phi}\bra{\phi})$, we have that the probability density of getting the
pointer reading $x$ is given by
\begin{eqnarray}\label{sec2:prob-s1}
p_\theta(x)=\tr_{\mathscr{SA}}\left[\ket{x}\bra{x}\mathcal{E}_T(\rho_\theta
\otimes
\ket{\phi}\bra{\phi})\right].
\end{eqnarray}
Besides, as shown in the main text, there is
\begin{eqnarray}
\mathcal{E}_T(\rho_\theta\otimes\ket{p}\bra{p^\prime})
=\left(e^{\mathcal{L}_{p,p^\prime}T}\rho_\theta\right)\otimes\ket{p}\bra{p^\prime},
\end{eqnarray}
where
\begin{eqnarray}
\mathcal{L}_{p,p^\prime}\rho=\mathcal{L}_\theta\rho-
iT^{-1}\left(pA\rho-p^\prime \rho A\right).
\end{eqnarray}
Expressing $\ket{\phi}$ in Eq.~(\ref{sec2:prob-s1}) as $\ket{\phi}=\int dp\ \phi(p)
\ket{p}$ with
\begin{eqnarray}
\phi(p)=\frac{1}{(2\pi\sigma^{\prime 2})^{1/4}}e^{-\frac{p^2}{4\sigma^{\prime2}}}
\end{eqnarray}
being the momentum representation of $\ket{\phi}$, where $\sigma^\prime=1/(2\sigma)$,
we have
\begin{eqnarray}\label{sec2:prob2}
p_\theta(x)=\frac{1}{2\pi}\int dpdp^\prime\,\phi(p)\phi^*(p^\prime)
e^{i(p-p^\prime)x}\tr_\mathscr{S}\left[e^{\mathcal{L}_{p,p^\prime}T}\rho_\theta\right].
\nonumber\\
\end{eqnarray}
Note that any linear map $\Lambda$ defined over the operator space of $\mathscr{S}$
has a dual map $\Lambda^*$, which is the map such that
\begin{eqnarray}\label{sec2:df:dual}
\tr_\mathscr{S}[X^\dagger\Lambda(Y)]=\tr_\mathscr{S}[\Lambda^*(X)^\dagger Y],
\end{eqnarray}
where $X$ and $Y$ are two linear operators acting on the Hilbert space of
$\mathscr{S}$. To digest the above definition, one can rewrite
Eq.~(\ref{sec2:df:dual}) as
$\expt{X,\Lambda(Y)}=\expt{\Lambda^*(X),Y}$,
where $\expt{X,Y}:=\tr_\mathscr{S}[X^\dagger Y]$ is known as the  Hilbert-Schmidt
inner product \cite{2010Nielsen}. So, the dual map $\Lambda^*$ of a map $\Lambda$
is defined in a way that is completely analogous to how the Hermitian conjugate
$X^\dagger$ of a linear operator $X$ is defined. It is not difficult to see that
the dual map of $e^{\mathcal{L}_{p,p^\prime}T}$ is
$e^{\mathcal{L}_{p,p^\prime}^*T}$ with
\begin{widetext}
	\begin{eqnarray}\label{sec2:dual-L}
	\mathcal{L}_{p,p^\prime}^*X=\gamma\left[
	\theta\left(\sigma_{+}X\sigma_{-}-\frac{1}{2}\{\sigma_{+}
	\sigma_{-},X\}\right)
	+(1-\theta)\left(\sigma_{-}X\sigma_{+}-\frac{1}{2}\{\sigma_{-}
	\sigma_{+},X\}\right)\right]+iT^{-1}\left(pAX-p^\prime XA\right).
	\end{eqnarray}
	With the above knowledge, we can rewrite the term
	$\tr_\mathscr{S}[e^{\mathcal{L}_{p,p^\prime}T}\rho_\theta]$
	appearing in Eq.~(\ref{sec2:prob2}) as
	$\tr_\mathscr{S}[(e^{\mathcal{L}_{p,p^\prime}^*T}I
	)^\dagger
	\rho_\theta]$. From this fact and Eq.~(\ref{sec2:prob2}), it follows that
	\begin{eqnarray}\label{sec2:prob3}
	p_\theta(x)=\tr_\mathscr{S}\left[\frac{1}{2\pi}\int dpdp^\prime\,\phi(p)\phi^*(p^\prime)
	e^{i(p-p^\prime)x}\left(e^{\mathcal{L}_{p,p^\prime}^*T}I\right)^\dagger
	\rho_\theta\right].
	\end{eqnarray}
	Comparing Eq.~(\ref{sec2:prob3}) with Eq.~(\ref{df:p}), we arrive at the
	expression of $\Pi_x$,
	\begin{eqnarray}\label{sec2:pi-exp}
	\Pi_x=\frac{1}{2\pi}\int dpdp^\prime\,\phi(p)\phi^*(p^\prime)
	e^{i(p-p^\prime)x}\left(e^{\mathcal{L}_{p,p^\prime}^*T}I\right)^\dagger.
	\end{eqnarray}
	As can be seen from Eq.~(\ref{sec2:dual-L}), the term $\mathcal{L}_{p,p^\prime}^*$
	appearing in Eq.~(\ref{sec2:pi-exp}) is $\theta$-dependent. Hence,
	it is expected that $\Pi_x$ is $\theta$-dependent, too.
	
	To confirm that $\Pi_x$ is indeed $\theta$-dependent, let us take a closer look at
	Eq.~(\ref{sec2:pi-exp}) in the following. For concreteness, we set the measured
	observable to be $A_\textrm{opt}=\ket{0}\bra{0}$. Using \textit{Mathematica}, we can find out
	the explicit expression of $(e^{\mathcal{L}_{p,p^\prime}^*T}I)^\dagger$ appearing
	in Eq.~(\ref{sec2:pi-exp}),
	\begin{eqnarray}\label{sec2:dual-L-exp}
	\left(e^{\mathcal{L}_{p,p^\prime}^*T}I\right)^\dagger=
	e^{-\left(T\gamma+i\nu\right)/{2}}
	\begin{pmatrix}
	\cosh(\mathfrak{S}/{2})+(T\gamma-i\nu)\frac{\sinh(\mathfrak{S}/2)}
	{\mathfrak{S}} & 0\\
	0 & \cosh(\mathfrak{S}/{2})+(T\gamma+i\nu)\frac{\sinh(\mathfrak{S}/2)}
	{\mathfrak{S}} \\
	\end{pmatrix},
	\end{eqnarray}
\end{widetext}
with
\begin{eqnarray}
\mathfrak{S}=\sqrt{T^2\gamma^2+2iT\gamma\nu(1-2\theta)-\nu^2}.
\end{eqnarray}
Here, for ease of notation, we have introduced the new variables
\begin{eqnarray}
\mu=p+p^\prime,\quad
\nu=p-p^\prime.
\end{eqnarray}
Substituting Eq.~(\ref{sec2:dual-L-exp}) into Eq.~(\ref{sec2:pi-exp}) and
noting that $\phi(p)\phi^*(p^\prime)=\frac{1}{(2\pi\sigma^{\prime 2})^{1/2}}
\exp[-\frac{\mu^2+\nu^2}{8\sigma^{\prime 2}}]$ and $d\mu d\nu=2dpdp^\prime$, we have
\begin{eqnarray}\label{sec2:Pi-new}
\Pi_x=\int f(\mu)d\mu
\begin{pmatrix}
\int g_1(\nu)d\nu & 0\\
0 & \int g_2(\nu)d\nu \\
\end{pmatrix}.
\end{eqnarray}
Here,
\begin{eqnarray}
f(\mu)&:=&\frac{1}{4\pi}\frac{1}{(2\pi\sigma^{\prime 2})^{1/2}}
\exp\left(-\frac{\mu^2}{8\sigma^{\prime 2}}\right),\\
g_1(\nu)&:=&\exp\left(-\frac{\nu^2}{8\sigma^{\prime 2}}+i\nu x-
\frac{T\gamma+i\nu}{2}\right)\times\nonumber\\
&&\left[\cosh(\mathfrak{S}/{2})+
(T\gamma-i\nu)\frac{\sinh(\mathfrak{S}/2)}{\mathfrak{S}}\right],\label{g1}\\
g_2(\nu)&:=&\exp\left(-\frac{\nu^2}{8\sigma^{\prime 2}}+i\nu x-
\frac{T\gamma+i\nu}{2}\right)\times\nonumber\\
&&\left[\cosh(\mathfrak{S}/{2})+
(T\gamma+i\nu)\frac{\sinh(\mathfrak{S}/2)}{\mathfrak{S}}\right].\label{g2}
\end{eqnarray}
While it is easy to perform the integration $\int f(\mu)d\mu$, i.e.,
$\int f(\mu)d\mu=\frac{1}{2\pi}$,
it is quite difficult to analytically work out the integration
$\int g_i(\nu)d\nu$, $i=1,2$, in general. To bypass this difficulty,
we consider the two limiting cases of $T\rightarrow 0$ and $T\rightarrow\infty$.

Consider first the limiting case of $T\rightarrow 0$. Using Taylor-series expansions
for $g_1(\nu)$ and $g_2(\nu)$ about $T=0$, we have, up to first order,
\begin{eqnarray}\label{sec2:g1-0}
g_1(\nu)&=&e^{i(x-1)\nu}e^{-\frac{\nu^2}{8\sigma^{\prime 2}}}-T\gamma(1-\theta)
\times\nonumber\\
&&e^{i(x-1)\nu}e^{-\frac{\nu^2}{8\sigma^{\prime 2}}}\left[1-\sum_{n=1}^\infty
\frac{(i\nu)^{n-1}}{n!}\right],
\end{eqnarray}
and
\begin{eqnarray}\label{sec2:g2-0}
g_2(\nu)&=&e^{ix\nu}e^{-\frac{\nu^2}{8\sigma^{\prime 2}}}-
T\gamma\theta\times\nonumber\\
&&e^{i x\nu}e^{-\frac{\nu^2}{8\sigma^{\prime 2}}}\left[1-\sum_{n=1}^\infty
\frac{(-i\nu)^{n-1}}{n!}\right].
\end{eqnarray}
Substituting Eqs.~(\ref{sec2:g1-0}) and (\ref{sec2:g2-0}) into
Eq.~(\ref{sec2:Pi-new}),
we obtain, after some algebra,
\begin{eqnarray}\label{sec2:Pi-two-terms}
\Pi_x=\Pi_x^{(0)}+T\Pi_x^{(1)},
\end{eqnarray}
with
\begin{eqnarray}\label{sec2:Pi-term-0}
\Pi_x^{(0)}=
\begin{pmatrix}
\frac{1}{\sqrt{2\pi\sigma^2}}e^{-\frac{(x-1)^2}{2\sigma^2}} & 0 \\
0 & \frac{1}{\sqrt{2\pi\sigma^2}}e^{-\frac{x^2}{2\sigma^2}}
\end{pmatrix}
\end{eqnarray}
and
\begin{widetext}
	\begin{eqnarray}\label{sec2:Pi-term-1}
	\Pi_x^{(1)}=
	\begin{pmatrix}
	-\gamma(1-\theta)\left[\frac{1}{\sqrt{2\pi\sigma^2}}e^{-\frac{(x-1)^2}{2\sigma^2}}
	-\frac{\textrm{erf}\left(\frac{x}{\sqrt{2}\sigma}\right)-
		\textrm{erf}\left(\frac{x-1}{\sqrt{2}\sigma}\right)}{2}\right] & 0 \\
	0 & -\gamma\theta\left[\frac{1}{\sqrt{2\pi\sigma^2}}e^{-\frac{x^2}{2\sigma^2}}
	-\frac{\textrm{erf}\left(\frac{x}{\sqrt{2}\sigma}\right)-
		\textrm{erf}\left(\frac{x-1}{\sqrt{2}\sigma}\right)}{2}\right]
	\end{pmatrix}.
	\end{eqnarray}
\end{widetext}
Here, $\textrm{erf}(x)$ is known as the Gauss error function,
defined as $\textrm{erf}(x)=\frac{1}{\sqrt{\pi}}\int_{-x}^{x}e^{-t^2}dt$.
Evidently, $\Pi_x^{(0)}$ is $\theta$-independent but $\Pi_x^{(1)}$ is
$\theta$-dependent.

In the limit of $T\rightarrow 0$, which corresponds to the PM,
we have that
\begin{eqnarray}
\Pi_x=\Pi_x^{(0)},
\end{eqnarray}
thereby reaching the textbook physics that the POVM operator is $\theta$-independent.
This can be understood on an intuitive level. Note that the interaction term
$H_I$ in Eq.~(\ref{sm-eq:measuring}) is unrelated to $\theta$, whereas the
dissipative term $\mathcal{L}_\theta$ is related to $\theta$ and has the
effect of making $\Pi_x$ $\theta$-dependent. In the limit of $T\rightarrow 0$,
the coupling strength is infinitely strong and the coupling time is infinitely short,
indicating that $H_I$ dominates Eq.~(\ref{sm-eq:measuring})
whereas $\mathcal{L}_\theta$ can be omitted from Eq.~(\ref{sm-eq:measuring}).
As a result, $\Pi_x=\Pi_x^{(0)}$ is, of course, $\theta$-independent.
However, in reality, any interaction is of finite strength and lasts for a
finite time interval, implying that $\mathcal{L}_\theta$ cannot be completely
ignored and plays some role in the measurement. So, in addition to $\Pi_x^{(0)}$,
the (small) $\theta$-dependent term $T\Pi_x^{(1)}$ resulting from the effect
of $\mathcal{L}_\theta$ appears in Eq.~(\ref{sec2:Pi-two-terms}).
This leads to the fact that $\Pi_x$ is $\theta$-dependent in practice.
\textit{Therefore, strictly speaking, the $\theta$-independence assumption may
	not hold even for PMs in practice.}

What happens if we enlarge $T$? As $H_I$ becomes weaker
as $T$ increases, $\mathcal{L}_\theta$ plays an increasingly important
role in the measurement. So, it can be expected that $\Pi_x$ depends
on $\theta$ increasingly heavily as $T$ increases. Roughly speaking,
the degree of the $\theta$-dependence of $\Pi_x$ achieves its maximum
in the limit of $T\rightarrow \infty$. \textit{Motivated by this,
	we let $T\rightarrow \infty$ in our proposal of measurements and expect that
	$\Pi_x$ depends on $\theta$ so heavily that the QCRB can be beaten by our measurement.}
This is the basic idea underlying our proposal. By the way,
as a matter of fact, the QCRB can be (slightly) beaten even by PMs with small but nonzero $T$, as shown below.
Let us now figure out the expression of $\Pi_x$ in the limiting case
of $T\rightarrow \infty$. It is not difficult to see that
\begin{eqnarray}\label{sec2:g12}
g_1(\nu)=g_2(\nu)=e^{i(x-\theta)\nu}e^{-\frac{\nu^2}{8\sigma^{\prime 2}}},
\end{eqnarray}
in this limit. Substituting Eq.~(\ref{sec2:g12}) into Eq.~(\ref{sec2:Pi-new}),
we have
\begin{eqnarray}\label{sec2:Pi-DAM}
\Pi_x=
\begin{pmatrix}
\frac{1}{\sqrt{2\pi\sigma^2}}e^{-\frac{(x-\theta)^2}{2\sigma^2}} & 0 \\
0 & \frac{1}{\sqrt{2\pi\sigma^2}}e^{-\frac{(x-\theta)^2}{2\sigma^2}}
\end{pmatrix},
\end{eqnarray}
in the limit of $T\rightarrow \infty$. Evidently,
$\Pi_x$ in Eq~(\ref{sec2:Pi-DAM}) is $\theta$-dependent.
As an immediate consequence, the $\theta$-independence assumption does not hold
for our proposal of measurements.

\subsection{Projective measurement: $T\rightarrow 0$}

We now inspect more carefully the limiting case of $T\rightarrow 0$.
It corresponds to the case that $H_I$ is extremely strong
but lasts for a very short time interval, i.e., $H_I$ is impulsive. Hereafter,
to make our discussion conceptually clear, we refer to the limit of $T\rightarrow 0$
as the impulsive limit. In contrast, the limit of $T\rightarrow\infty$ is
referred to as the adiabatic limit hereafter, which corresponds to DAMs to be discussed later on. In the impulsive limit, the
probability density of getting the pointer reading $x$ is given by
\begin{eqnarray}\label{sec2:PM-prob-fun}
p_\theta(x)=\frac{\theta}{\sqrt{2\pi\sigma^2}}e^{-\frac{(x-1)^2}{2\sigma^2}}+
\frac{1-\theta}{\sqrt{2\pi\sigma^2}}e^{-\frac{x^2}{2\sigma^2}},
\end{eqnarray}
which can be obtained by substituting Eq.~(\ref{sec2:Pi-term-0})
into Eq.~(\ref{df:p}).
\begin{figure}[htbp]
	\includegraphics[width=0.4\textwidth]{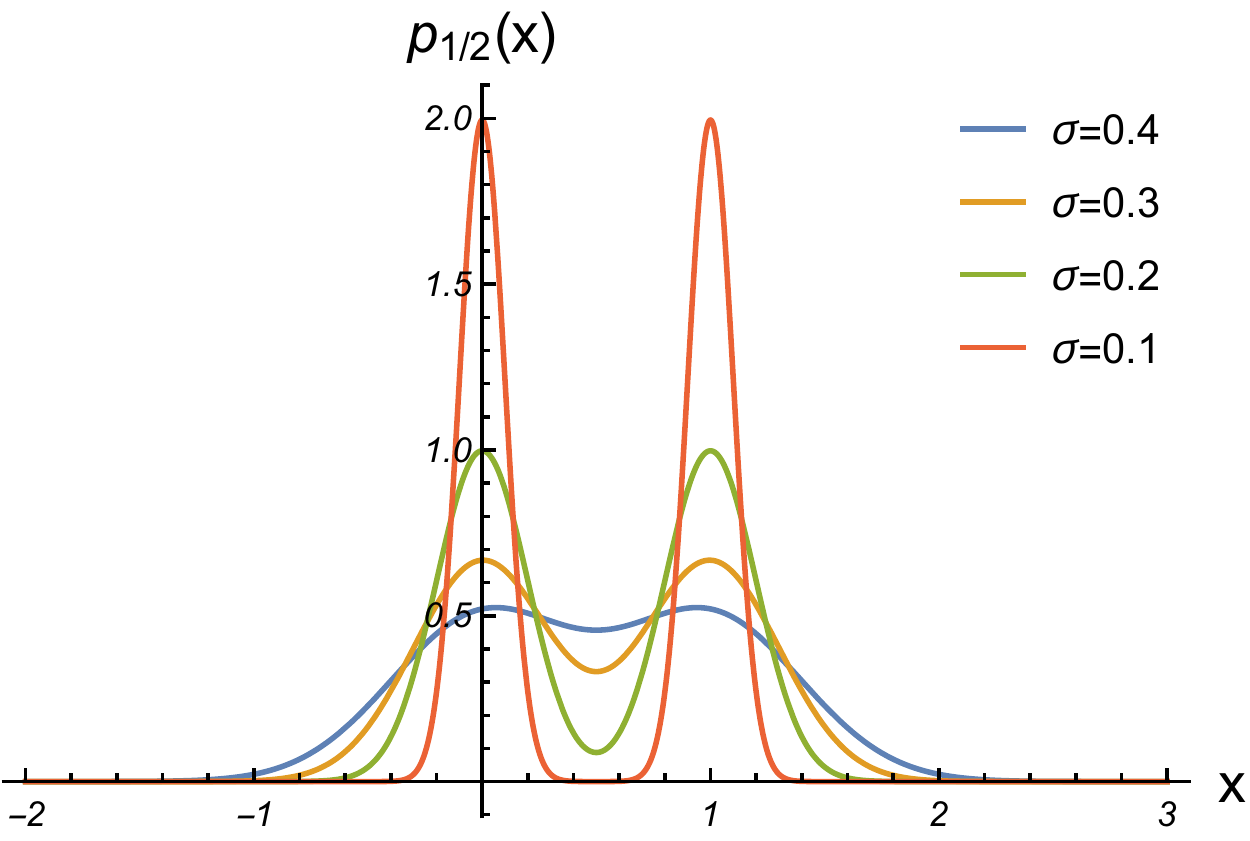}
	\caption{Probability density $p_{1/2}(x)$ of getting the pointer
		reading $x$ in the impulsive
		limit with different standard deviations $\sigma$.}
	\label{fig-PM-prob-fun}
\end{figure}
For the reader's information, we plot in Fig.~\ref{fig-PM-prob-fun} the
probability distribution $p_{1/2}(x)$ with different standard deviations
$\sigma$. As can be seen from Fig.~\ref{fig-PM-prob-fun} as well as
Eq.~(\ref{sec2:PM-prob-fun}), for a small $\sigma$, say,
$\sigma=0.1$, $p_{\theta}(x)$ is only significantly different from zero
if $x\approx 0$ or $x\approx 1$. Noting that $0$ and $1$ are the two eigenvalues
of $A_\textrm{opt}=\ket{0}\bra{0}$, we deduce that, roughly speaking, the pointer reading $x$,
as a random variable, is mostly likely to be one of these two eigenvalues in
the PM with a small $\sigma$. This is just the well-known
textbook physics that the potential outcome of a PM is one
of the eigenvalues of the measured observable. Strictly speaking, this textbook
physics is valid only in the mathematical limit of $\sigma\rightarrow 0$, for
which the continuous probability distribution $p_\theta(x)$ can be treated as a
discrete probability distribution. To put it differently, in practice,
where $\sigma$ is small but non-zero, the pointer reading $x$ may not be exactly
one of the eigenvalues; there could be small fluctuations due to the 	
uncertainty	of initial position of the pointer.

\begin{figure}[htbp]
	\includegraphics[width=0.4\textwidth]{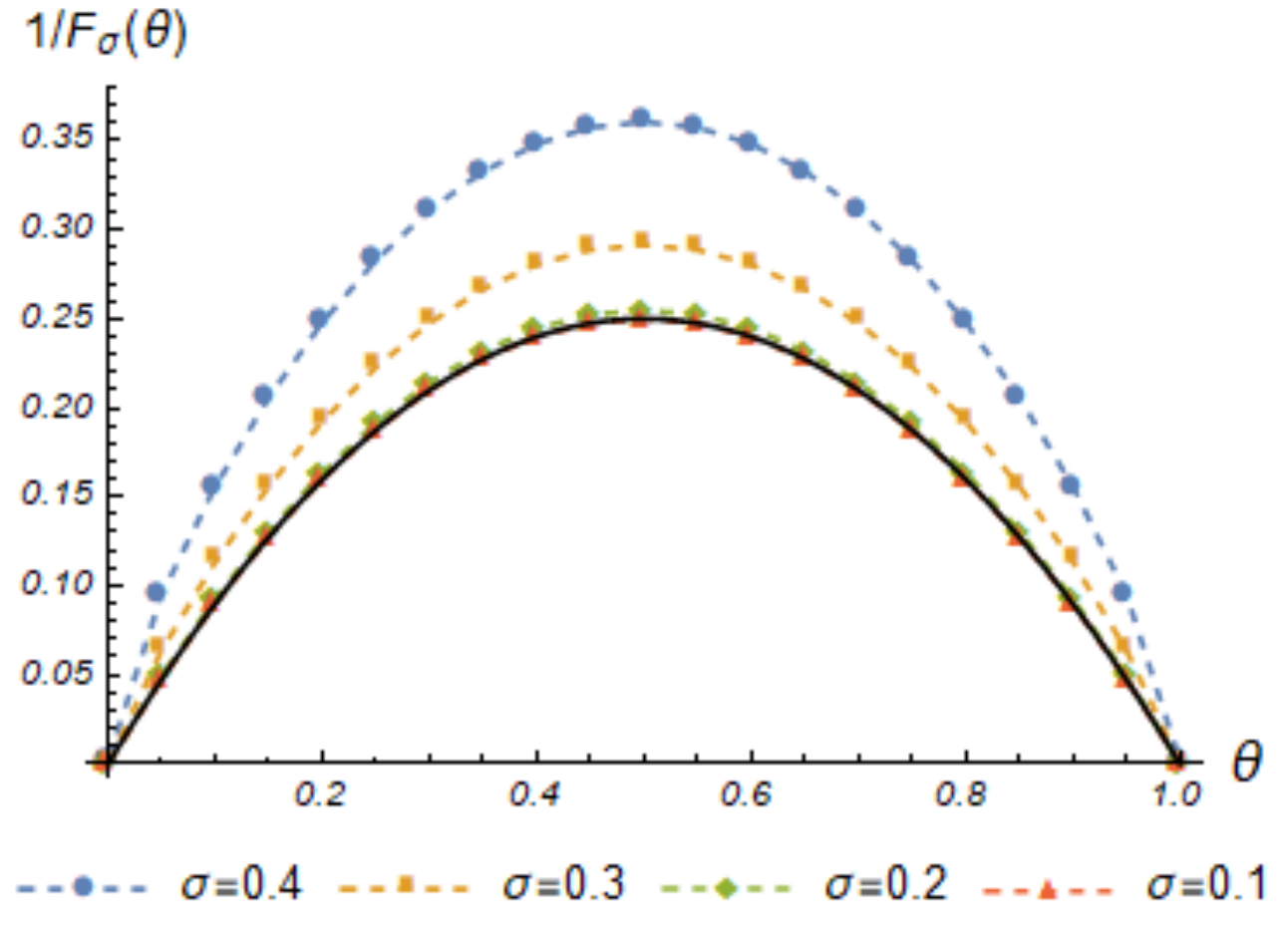}
	\caption{Numerical results of the classical Fisher information
		$F_\sigma(\theta)$ associated with $p_\theta(x)$ in Eq.~(\ref{sec2:PM-prob-fun})
		for four different $\sigma$.  The abscissa and the ordinate are $\theta$ and
		$1/F_\sigma(\theta)$, respectively. The black solid curve represents
		$1/H(\theta)$. For a given $\theta$, $1/F_\sigma(\theta)$
		gradually decreases as $\sigma$ decreases and finally approaches the
		limiting position specified by $1/H(\theta)$ in the limit of
		$\sigma\rightarrow 0$.}
	\label{fig-PM-FI}
\end{figure}

It is easy to see that the QFI associated with $\rho_\theta$ is given by
\begin{eqnarray}\label{sec2:PM-QFI}
H(\theta)=\frac{1}{\theta(1-\theta)}.
\end{eqnarray}
Equation (\ref{sec2:PM-QFI}) can be obtained by first solving Eq.~(\ref{SLD})
to get $L_\theta=\textrm{diag}(\frac{1}{\theta},-\frac{1}{1-\theta})$ and then
inserting $L_\theta$ into Eq.~(\ref{CFI-QFI}).
Denote by $F_\sigma(\theta)$ the CFI associated with $p_\theta(x)$ in
Eq.~(\ref{sec2:PM-prob-fun}). Here, the subscript $\sigma$ is used to indicate that
the CFI is dependent of $\theta$ because of the $\theta$-dependence of $p_\theta(x)$.
Noting that the $\theta$-independence assumption and, therefore, the QCRB \GJ{holds}
in the impulsive limit, we have
\begin{eqnarray}\label{sec2:PM-CFI-QFI}
F_\sigma(\theta)\leq H(\theta).
\end{eqnarray}
On the other hand, it is not difficult to see that
\begin{eqnarray}\label{sec2:PM-CFI-limit}
\lim_{\sigma\rightarrow 0}F_\sigma(\theta)=H(\theta).
\end{eqnarray}
Indeed, in the limit of $\sigma\rightarrow 0$, $p_\theta(x)$ in
Eq.~(\ref{sec2:PM-prob-fun}) can be treated as the discrete probability
distribution with the probability of getting $1$ being $\theta$ and that of
getting $0$ being $1-\theta$. Using Eq.~(\ref{CFI}), one can confirm that the
CFI associated with this discrete probability distribution is exactly the QFI
given by Eq.~(\ref{sec2:PM-QFI}). Note that the observable corresponding
to $p_\theta(x)$ in Eq.~(\ref{sec2:PM-prob-fun}) is  $A_\textrm{opt}=\ket{0}\bra{0}$.
The above point indicates that $A_\textrm{opt}=\ket{0}\bra{0}$ is optimal, since the QCRB
can be achieved if one performs the PM associated with \GJ{$A_\textrm{opt}$}.
Combing Eqs.~(\ref{sec2:PM-CFI-QFI}) and (\ref{sec2:PM-CFI-limit}), we have
\begin{eqnarray}
F_\sigma(\theta)\leq \lim_{\sigma\rightarrow 0}F_\sigma(\theta)=H(\theta).
\end{eqnarray}
Figure \ref{fig-PM-FI} shows the numerical results of $F_\sigma(\theta)$
for four different $\sigma$, namely, $\sigma=0.4,0.3,0.2,0.1$.
As can be seen from Fig.~\ref{fig-PM-FI}, $1/{F_\sigma(\theta)}$ is strictly
larger than $1/{H(\theta)}$ for $\sigma>0$ but approximately equals
to $1/{H(\theta)}$ for a small enough $\sigma$, say, $\sigma=0.2$.
Therefore, to achieve the QCRB, in addition to choosing an optimal observable,
one needs to prepare the measuring apparatus $\mathscr{A}$ in a state with
a small $\sigma$.

\begin{figure}
	\centering
	\subfigure[]{
		\includegraphics[width=.4\textwidth]{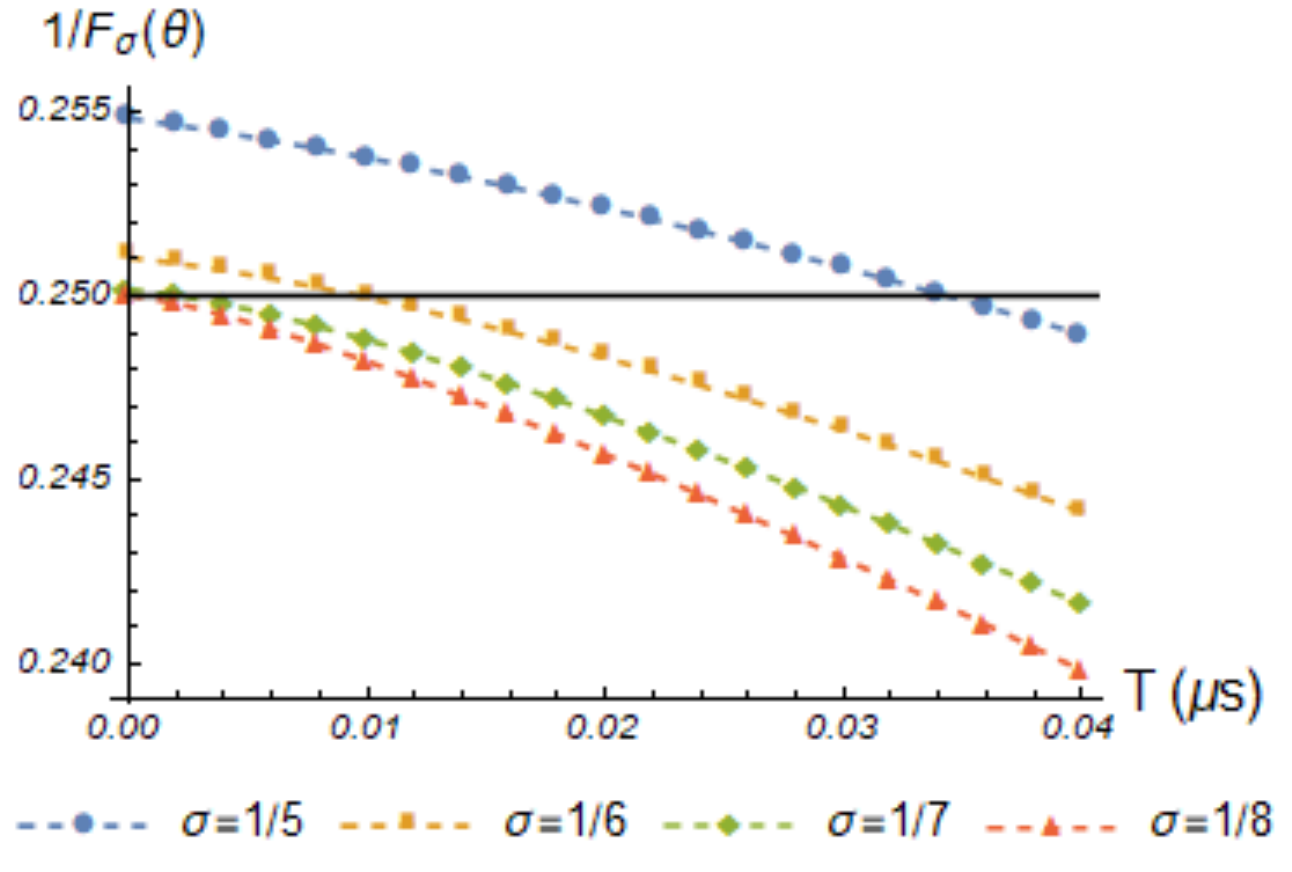}\label{fig-FI-small-T-5}}
	\subfigure[]{
		\includegraphics[width=.4\textwidth]{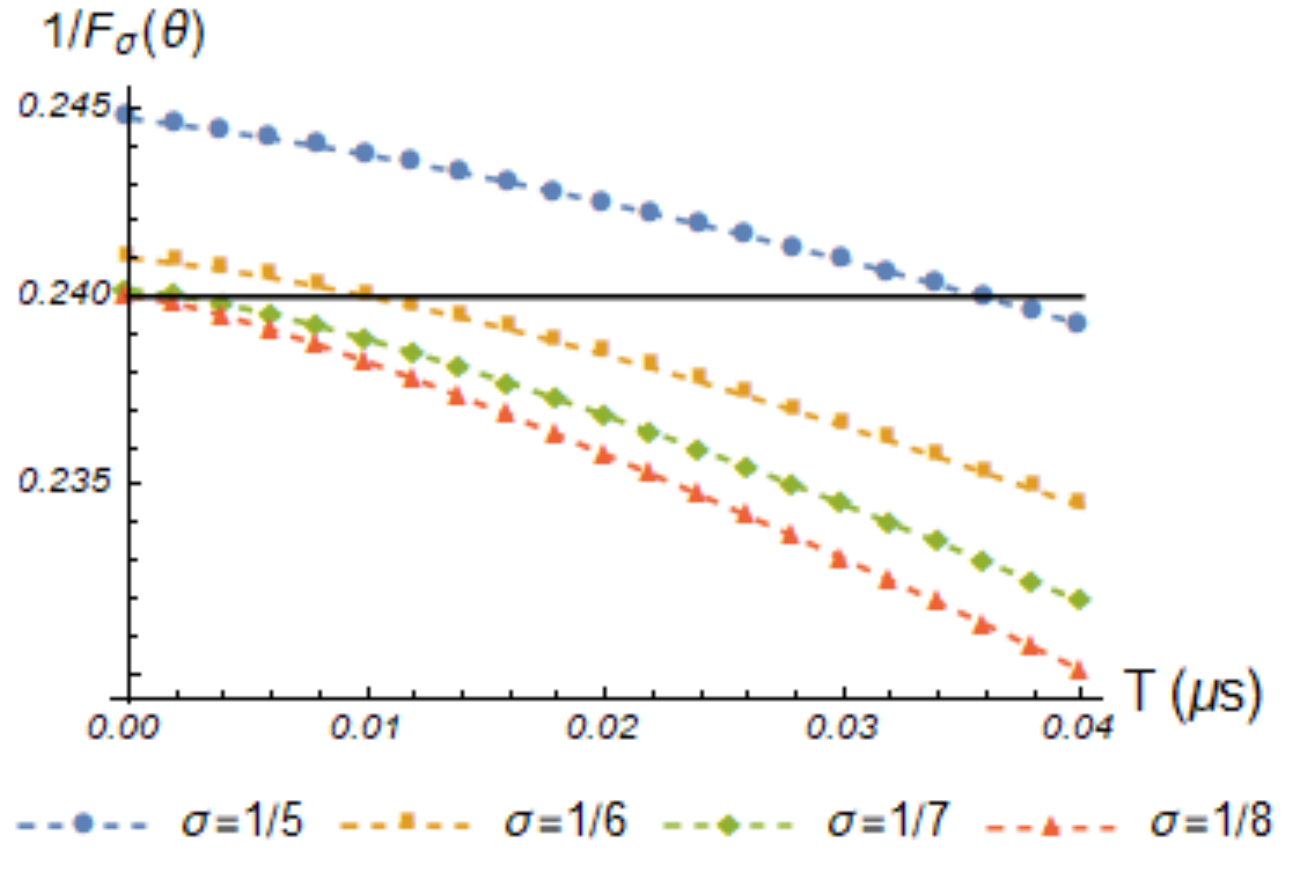}\label{fig-FI-small-T-4}}
	\subfigure[]{
		\includegraphics[width=.4\textwidth]{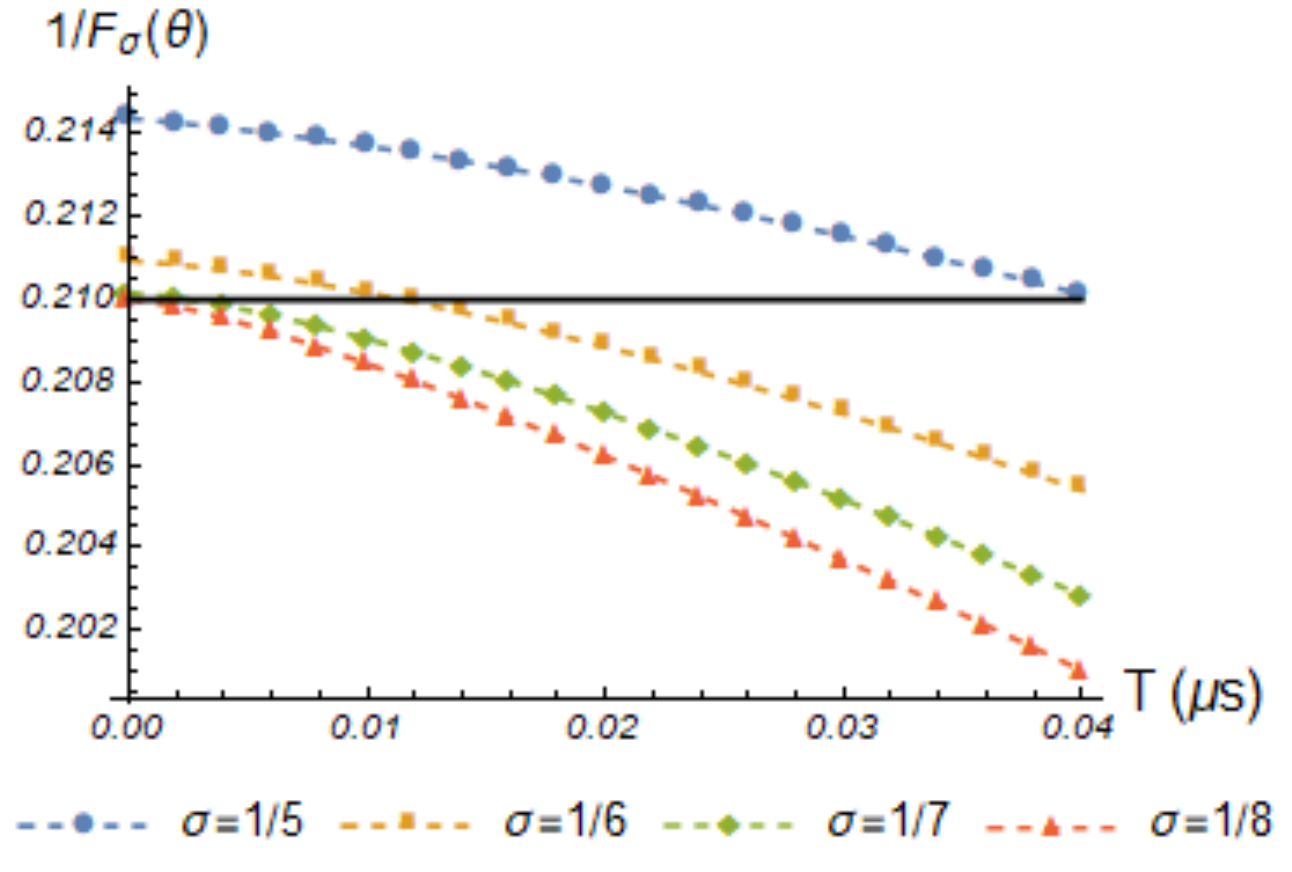}\label{fig-FI-small-T-3}}
	\caption{Numerical results of $1/{F_\sigma(\theta)}$ with small values of $T$ for
		three different $\theta$ and four different $\sigma$. (a) $\theta=0.5$. (b)
		$\theta=0.4$. (c)
		$\theta=0.3$. The three black solid lines in
		Figs.~\ref{fig-FI-small-T-5}-\ref{fig-FI-small-T-3}
		represent the
		quantum Cram\'{e}r-Rao bounds
		$1/{H(\theta)}$ associated with these three $\theta$, respectively. Here,
		$\gamma$ is set to be $5$ MHz.}
	\label{fig-FI-small-T}
\end{figure}

Consider now the PM with
a small but nonzero $T$, for which the $\theta$-independence assumption does
not hold.
Substituting Eq.~(\ref{sec2:Pi-two-terms}) into Eq.~(\ref{df:p}),
we have that the probability density of getting the pointer reading $x$
is given by
\begin{eqnarray}\label{sec2:pm-prob-fun-small-T}
&&p_\theta(x)=\frac{\theta-T\gamma\theta(1-\theta)}{\sqrt{2\pi\sigma^2}}
e^{-\frac{(x-1)^2}{2\sigma^2}}+
\frac{1-\theta-T\gamma\theta(1-\theta)}{\sqrt{2\pi\sigma^2}}\times\nonumber\\
&&e^{-\frac{x^2}{2\sigma^2}}+T\gamma\theta(1-\theta)\left[\textrm{erf}(
\frac{x}{\sqrt{2}\sigma})-\textrm{erf}(
\frac{x-1}{\sqrt{2}\sigma})
\right].
\end{eqnarray}
To examine whether the
QCRB can be beaten in this case, we numerically compute $1/{F_\sigma(\theta)}$
associated with the $p_\theta(x)$ in Eq.~(\ref{sec2:pm-prob-fun-small-T}) for three
different $\theta$ and four different $\sigma$,
namely, $\theta=0.5,0.4,0.3$ and $\sigma=1/5,1/6,1/7,1/8$. The numerical results are
shown in Fig.~\ref{fig-FI-small-T}, where we set $\gamma=5$ MHz.
The three black solid lines depicted
in Figs.~\ref{fig-FI-small-T-5}-\ref{fig-FI-small-T-3} represent the QCRBs
$1/{H(\theta)}$ associated with these three values of
$\theta$, respectively. As can be
seen
from Fig.~\ref{fig-FI-small-T}, $1/{F_\sigma(\theta)}$ can be strictly
less than
$1/{H(\theta)}$ for some small but nonzero $T$, indicating that
the QCRB can be beaten for the PM
with a small but nonzero $T$, as claimed in the previous paragraph.

\subsection{Dissipative adiabatic measurement: $T\rightarrow\infty$}

\begin{figure}[htbp]
	\includegraphics[width=0.4\textwidth]{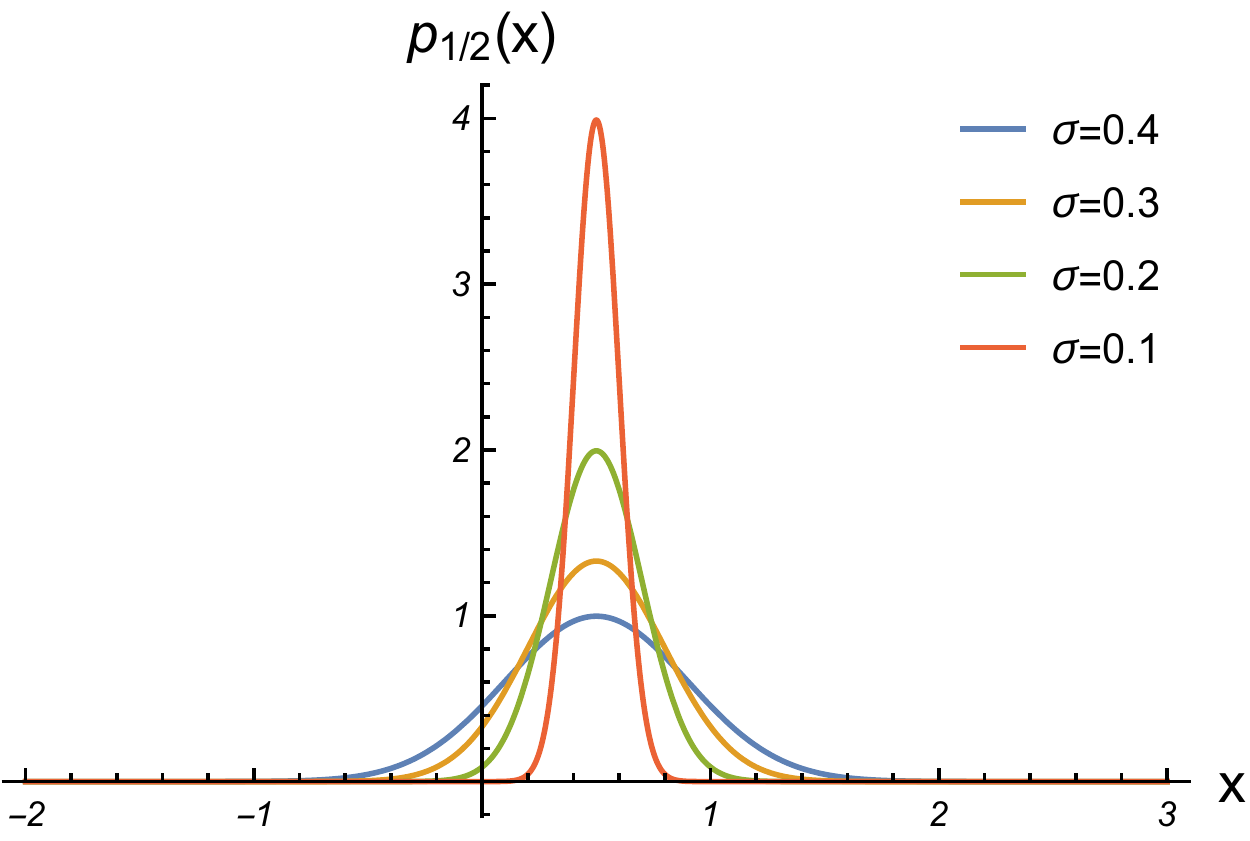}
	\caption{Probability density $p_{1/2}(x)$ of
		getting the pointer reading $x$ in the adiabatic limit with
		different standard deviations $\sigma$.}
	\label{fig-DAM-prob-fun}
\end{figure}

Consider now the adiabatic limit. In this limit, the
probability density of getting outcome $x$ reads
\begin{eqnarray}\label{sec2:DAM-prob-fun}
p_\theta(x)=\frac{1}{\sqrt{2\pi\sigma^2}}e^{-\frac{(x-\theta)^2}{2\sigma^2}},
\end{eqnarray}
which can be obtained by substituting Eq.~(\ref{sec2:Pi-DAM})
into Eq.~(\ref{df:p}).
For the reader's information, we plot in Fig.~\ref{fig-DAM-prob-fun} the
probability distribution $p_{1/2}(x)$ with different standard deviations
$\sigma$.
As can be seen from Fig.~\ref{fig-DAM-prob-fun} as well as
Eq.~(\ref{sec2:DAM-prob-fun}), for a small $\sigma$, say,
$\sigma=0.1$, $p_{\theta}(x)$ is only significantly different from zero
if $x\approx \theta$. Note that $\theta$ is the expectation value of $A_\textrm{opt}$ in the
state $\rho_\theta$. So, roughly speaking,
the outcome of the DAM
is the expectation value of the measured observable, which is in sharp contrast
\GJ{to} that of the PM, i.e., an individual eigenvalue of
the measured observable.

\begin{figure}[htbp]
	\includegraphics[width=0.4\textwidth]{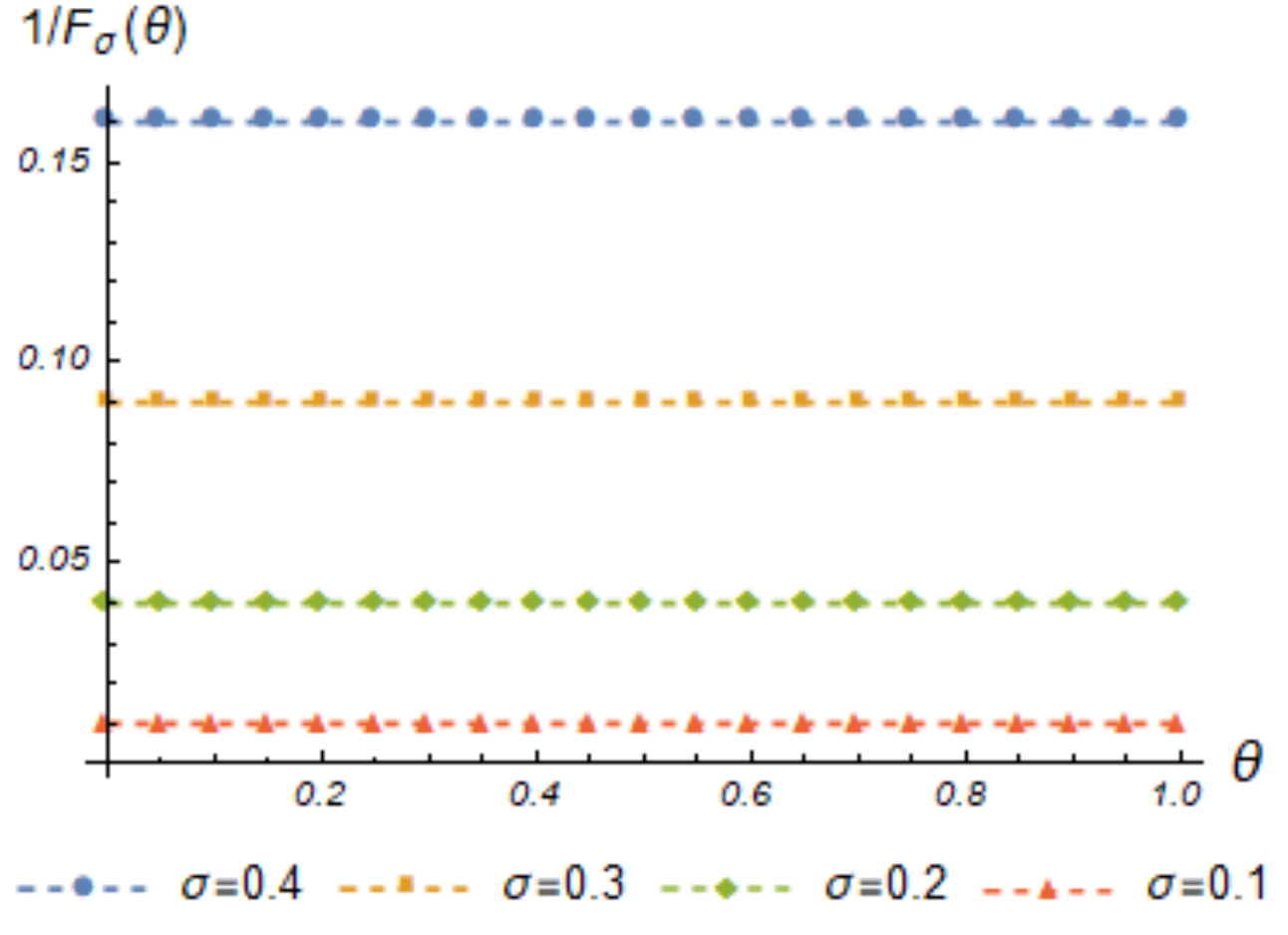}
	\caption{Classical Fisher information
		$F_\sigma(\theta)$ associated with $p_\theta(x)$ in Eq.~(\ref{sec2:DAM-prob-fun})
		for four different $\sigma$.  The abscissa and the ordinate are $\theta$ and
		$1/{F_\sigma(\theta)}$, respectively.
		For a given $\theta$, $1/{F_\sigma(\theta)}$
		decreases as $\sigma$ decreases and finally approaches zero
		in the limit of
		$\sigma\rightarrow 0$.}
	\label{fig-DAM-FI}
\end{figure}

Substituting Eq.~(\ref{sec2:DAM-prob-fun}) into Eq.~(\ref{CFI}), we have that
the associated CFI reads
\begin{eqnarray}\label{sec2:DAM-CFI}
F_\sigma(\theta)=\frac{1}{\sigma^2}.
\end{eqnarray}
Figure \ref{fig-DAM-FI} shows the profiles of $1/{F_\sigma(\theta)}$ for
four different $\sigma$. As can be easily seen from Eq.~(\ref{sec2:DAM-CFI})
as well as Fig.~\ref{fig-DAM-FI}, $F_\sigma(\theta)$ can be arbitrarily large
so long as $\sigma$ is small enough. Therefore,
there is no intrinsic bound on precision in the DAM, \GJ{in sharp contrast to the ideal PM case} (see Fig.~\ref{fig-PM-FI}).
To better digest this point, one may proceed as follows. Suppose that
we are only given one data, which is obtained from the measurement associated with
$A_\textrm{opt}=\ket{0}\bra{0}$. If the measurement is a PM, this data,
denoted as $x^{\textrm{PM}}$, is either $1$ or $0$, provided that $\sigma$ is
very small. Evidently, the data $x^{\textrm{PM}}$ itself
is unrelated to the parameter $\theta$. Therefore, given only one data $x^\textrm{PM}$,
there is no way to
definitely determine $\theta$ from $x^\textrm{PM}$ even in principle. However,
if the measurement is a DAM, this data, denoted as
$x^\textrm{DAM}$, is approximately
equal to $\theta$. So, $x^{\textrm{DAM}}$
is more informative than $x^{\textrm{PM}}$, as $x^{\textrm{DAM}}$ is directly
related to $\theta$. Such a direct relationship allows one to extract
the value of $\theta$ from $x^\textrm{DAM}$ straightforwardly,
leading to the fact that $\theta$ can be determined to any degree of accuracy
so long as $\sigma$ is small enough.


\begin{thebibliography}{47}
\bibitem [{1no()}]{1note}%
	\BibitemOpen
	\href@noop {} {}\bibinfo {note} {Unless otherwise stated, the Hilbert-Schmidt
		norm is adopted. That is, for an operator $X$, the norm reads
		$\norm{X}:=\sqrt{\tr(X^\dagger X)}$, while for a superoperator $\mathcal{E}$,
		it is the induced norm defined as $\norm{\mathcal{E}}:=\sup_{\norm{X}\leq
			1}\norm{\mathcal{E}(X)}$.}\BibitemShut {Stop}%
\end{thebibliography}
%

\end{document}